\documentclass[10pt, nolongbibliography,nofootinbib,groupedaddress,showpacs,showkeys,amsmath,amssymb,eqsecnum,aps,prd]{revtex4-2}

\usepackage{microtype}

\usepackage{newtx}
\usepackage{booktabs}
\usepackage{tensor, mathtools, bm} 

\usepackage{epsfig}

\usepackage[dvipsnames]{xcolor}
\usepackage{graphicx} 
\usepackage[colorlinks, allcolors=Blue]{hyperref}

\begin{document}   
\title{Quantum gravity and matter fields in a general background gauge}
\begin{abstract}
    We analyse the gauge-dependence of the effective action in an interacting quantum theory of gravitational and matter fields. An explicit off-shell result is obtained in a general background gauge at one-loop order, which reduces in a particular gauge to the effective action found by 't Hooft-Veltman. We confirm the validity of DeWitt-Kallosh theorem, which implies that the on-shell effective action should be independent of the gauge-fixing parameter. We employ this theorem to expose the non-renormalizability of the theory in a general background gauge.
\end{abstract}

\author{J. Frenkel}
\email{jfrenkel@if.usp.br}
\affiliation{Instituto de F\'{\i}sica, Universidade de S\~ao Paulo, S\~ao Paulo, SP 05508-090, Brazil}

\author{S. Martins-Filho}   
\email{s.martins-filho@unesp.br}
\affiliation{Instituto de F\'{\i}sica Te\'orica, Universidade Estadual Paulista (UNESP), Rua Dr.\ Bento Teobaldo Ferraz 271 - Bloco II, 01140-070 S\~ao Paulo, SP, Brazil}

\date{\today}

\maketitle

\section{Introduction} \label{sec:intro} 

Classical general relativity is a successful theory which provides a very good description of the gravitational interactions that occur at low energies. There have been many attempts to quantize gravity along the lines used in other gauge theories and it was recognized that Einstein's theory of gravity is generally not renormalisable in the usual sense. The contributions generated by the Feynman loops to all orders require an infinite number of counterterms to cancel the ultraviolet divergences present in this theory. The current point of view is that general relativity could be considered as a low energy effective field theory of some fundamental theory \cite{weinberg:1995, Donoghue:1995cz}. 

Pure quantum gravity is on-shell renormalizable at one-loop order, so that its divergences can be absorbed by a redefinition of the fields. On the  other hand, it was shown in Ref. \cite{tHooft:1974toh, *thooft:1993}  that when a scalar field is coupled to gravity, the theory is no longer on-shell renormalizable at one-loop order, in the sense that its divergences cannot be removed by field redefinitions. This outcome has been found in a particular background-field gauge.

      Many of the perturbative computations done in quantum gravity used the background-field method, which preserves gauge invariance at all stages of the calculations \cite{tHooft:1974toh, Abbott:1980hw, Barvinsky:2017zlx, Frenkel:2018xup}. It was shown that the $S$-matrix calculated in this approach is equivalent to the conventional $S$-matrix \cite{Abbott:1983zw}. Thus, one expects the on-shell counterterms to be independent of the gauge-fixing parameters and of the parametrization of graviton fields. This statement was proved by DeWitt and Kallosh through a formal  theorem, which asserts that the loop counterterms on mass-shell should be independent of the gauge-fixing and of the parametrization \cite{DeWitt:1967ub, Kallosh:1974yh}. Their arguments have been  subsequently improved in several papers \cite{Grisaru:1975ei, Kallosh:1978wt, Fradkin:1981iu, deBerredo-Peixoto:2000vzo, Lavrov:1995kr, Lavrov:2019nuz}. 

However, some questions have been raised on the general validity of this theorem in the non-renormalisable theory of quantum gravity \cite{Ichinose:1991ap, *Ichinose:1992np, Falls:2015qga, Ohta:2016npm}. Nevertheless, the gauge-fixing independence of the on-shell effective action was explicitly checked, in a rather general gauge, at one-loop order in quantum gravity with a cosmological constant \cite{Kalmykov:1998cv}. 
Moreover, the on-shell independence of the effective action in this model has been verified for a very general parametrization of the metric field \cite{Goncalves:2017jxq}.

It may be useful to verify through explicit calculations the DeWitt–Kallosh theorem for quantum gravity in the presence of matter fields. A wide class of such theories has drawn considerable attention in the areas of quantum \cite{Barvinsky:1992tr, *Barvinsky:1995gi} and inflationary \cite{Spokoiny:1984bd, *Fakir:1990eg} cosmology, where the scalar fields are nonminimally coupled to gravity in a very involved way. 
For these reasons, we consider in this paper a simple coupled scalar-gravity theory. We use the background field method with a general nonminimal gauge-fixing Lagrangian\footnote{A systematic approach to handle nonminimal gauges in the Schwinger-DeWitt thecnique were derived in \cite{Barvinsky:1985an} for pure quantum gravity.}, that is characterized by two generic gauge parameters: $ \xi $ and $\zeta$. 
We will show that, in the limit $ \xi \to 0$, there appear in individual Feynman diagrams singularities which could hinder the general validity of this theorem. However, it turns out that such singularities cancel out when adding all diagrams.

We outline the Becchi-Rouet-Stora-Tyutin (BRST) \cite{Becchi:1975nq, *Tyutin:1975qk} quantization of this theory in Sec. II, where we clarify some aspects of the DeWitt-Kallosh theorem within the context of this theory. In Sec. III, we present the calculation of the off-shell effective action at one-loop order, which explicitly depends on the generic gauge parameters $ \xi $ and $\zeta$.
In particular, we recover the result found in Ref.~\cite{Grisaru:1975ei} in the limit $ \xi =1$, and reproduce ‘t Hooft-Veltman result in the limit $ \xi =\zeta=1$. We confirm this theorem, by showing  that the on-shell effective action at one-loop   is independent of the gauge parameters. As shown in Sec. IV, the DeWitt-Kallosh theorem enables to prove in a simple way that this theory remains non-renormalizable in a general background-field gauge. A summary of the results is presented in Sec. V. Some useful details of the calculations are given in the Appendices.    

\section{BRST symmetry and DeWitt-Kallosh theorem}

Let us consider, for definiteness, the interaction between a scalar field $ \bar{\phi} $ and a gravitational field $ \bar{g}^{\mu \nu} $ described by the Einstein-Hilbert Lagrangian \cite{tHooft:1974toh, *thooft:1993}
\begin{equation}\label{eq:21}
    \mathcal{L}^{\text{inv}} ( \bar{g} , \bar{\phi} ) = - \frac{1}{\kappa^{2}} \sqrt{ \bar{g}} \left [ \bar{R}  + \frac{\kappa^{2}}{2} \bar{g}^{\mu \nu} \partial_{\mu} \bar{\phi} \partial_{\nu} \bar{\phi}\right ],
\end{equation}
where $ \kappa^{2} = 16 \pi G_{N} $ and $ \bar{R} $ is the Ricci scalar.

In order to preserve the basic diffeomorphism symmetry of the theory, we use the background field method \cite{Abbott:1980hw}, where the fields are splitted into a background  and a quantum part as 
\begin{equation}\label{eq:22}
    \bar{g}_{\mu \nu} = g_{\mu \nu} + \kappa h_{\mu \nu} ; \quad \bar{\phi} = \phi +  \varphi.
\end{equation}
Here, $ g_{\mu \nu} $ and $ \phi $ are background fields, and $ h_{\mu \nu} $ and $ \varphi $ are quantum fields.
The Lagrangian \eqref{eq:21}, with the substitution \eqref{eq:22}, is invariant under two types of transformations.
The first type is a background-field  transformation given by 
\begin{subequations}\label{eq:23}
    \begin{align}\label{eq:23a}
        \frac{1}{\kappa } \bar{\delta} g_{\mu \nu} ={}& \epsilon^{\alpha} \partial_{\alpha} g_{\mu \nu} + g_{\mu \alpha} \partial_{\nu} \epsilon^{\alpha} + g_{\nu \alpha} \partial_{\mu} \epsilon^{\alpha} \equiv \tensor{E}{_{\mu\nu}_{\alpha}} ( g) \epsilon^{\alpha}, \qquad \frac{1}{\kappa }\bar{\delta} \phi = \epsilon^{\alpha} \partial_{\alpha} \phi; \\  
        \frac{1}{\kappa }\bar{\delta} h_{\mu \nu} ={}& \epsilon^{\alpha} \partial_{\alpha} h_{\mu \nu} + h_{\mu \alpha} \partial_{\nu} \epsilon^{\alpha} + h_{\nu \alpha} \partial_{\mu} \epsilon^{\alpha}, \qquad \frac{1}{\kappa }\bar{\delta} \varphi = \epsilon^{\alpha} \partial_{\alpha} \varphi.
    \end{align}
\end{subequations}
where $ \tensor{E}{^{\mu\nu}_{\alpha}} ( g)$ is covariant with respect to the background coordinate transformation.  

The second type is a quantum field transformation  given by 
\begin{subequations}\label{eq:24}
    \begin{align}\label{eq:24a}
        \delta g_{\mu \nu} ={}&0, \quad  \delta \phi =0; \\
         \delta h_{\mu \nu} ={}& \tensor{E}{_{\mu \nu}_{\alpha}} ( g_{\mu \nu} + \kappa h_{\mu \nu} ) \epsilon^{\alpha} ; \quad  \delta \varphi = \kappa \epsilon^{\alpha} \partial_{\alpha} ( \phi +  \varphi ). 
    \end{align}
\end{subequations}

In order to maintain the background symmetry  \eqref{eq:23}, but break the quantum symmetry \eqref{eq:24}, we introduce a gauge-fixing term given by 
\begin{equation}\label{eq:25}
    \mathcal{L}_{\text{gf}} ( g, \phi , h, \varphi ) = -\frac{1}{2 \xi} \sqrt{g} \left( h^{\mu \nu}_{; \nu} - \frac{1}{2} h_{\alpha}^{\alpha}{}^{; \mu} - \zeta \kappa
     \varphi \partial^{\mu} \phi    \right)^{2} 
\end{equation}
where $;$ denotes covariant derivatives with respect to the background field $ g_{\mu \nu} $ and $\xi$, $\zeta$ are two arbitrary gauge parameters.
The corresponding Faddeev-Popov ghost Lagrangian \cite{Faddeev:1967fc} is given by 
\begin{equation}\label{eq:27}
    \mathcal{L}_{\text{gh}} = \sqrt{g} \bar{\eta}^{ \mu} \left[ \tensor{\eta}{_{\mu ; \alpha}^{\alpha}} - R_{\mu \nu} \eta^{\nu} - \zeta \kappa \partial_{\mu} \phi \partial_{\nu} \phi \eta^{\nu} \right], 
\end{equation}
where $R_{\mu \nu} $ is the Ricci tensor, $ \bar{\eta}^{ \mu} $ and $ \eta^{\nu} $ are vector ghost fields, treated as quantum fields.
At this point, it turns out to be more convenient to use, instead of the gauge-fixing term \eqref{eq:25}, an equivalent Nakanishi-Lautrup Lagrangian \cite{Nakanishi:1966zz, lautrup:1967} defined as 
\begin{equation}\label{eq:28}
    \mathcal{L} '_{\text{gf} } =\sqrt{g} \left [ \frac{\xi}{2} B_{\mu} B^{\mu} + B_{\mu}\left( h^{\mu \nu}{}_{; \nu} - \frac{1}{2} h_{\alpha}^{\alpha}{}^{; \mu} - \zeta \kappa \varphi \partial^{\mu} \phi\right)      \right ],
\end{equation}
where $ B^{\mu} $ is a commuting quantum field. By functionally integrating over $B^{\mu} $, one can easily check the  equivalence of Eq.~\eqref{eq:28} to the gauge-fixing term Eq.~\eqref{eq:25}. 

Proceeding in this way, one can show that the total Lagrangian 
\begin{equation}\label{eq:29}
    \mathcal{L}_{\text{tot}} ( g , \phi , h, \varphi , B , \bar{\eta}^{} , \eta ) = \mathcal{L}^{\text{inv} } ( g + \kappa h, \phi +  \varphi ) + \mathcal{L} '_{\text{gf}} + \mathcal{L}_{\text{gh}}
\end{equation}
is invariant under the quantum BRST transformations \cite{Stelle:1976gc}
\begin{subequations}\label{eq:210}
    \begin{align}\label{eq:210a}
        & \delta g_{\mu \nu} = 0, \quad \delta \phi = 0; \\
        & \delta h_{\mu \nu} = \omega \tensor{E}{_{\mu \nu}_{\tau}} ( \bar{g} ) \eta^{\tau} \equiv \omega \mathop{\mathsf{s}} h_{\mu \nu} , \quad \delta \varphi = \kappa \omega ( \partial_{\tau} \bar{\phi} ) \eta^{\tau} \equiv \omega \mathop{\mathsf{s}}  \varphi; \\
        & \delta \eta^{\mu} = \kappa \omega \eta^{\nu} \eta^{\mu}_{; \nu} \equiv \omega \mathop{\mathsf{s}}  \eta^{\mu} ; \quad \delta \bar{\eta}^{}_{\mu} = - \omega B_{\mu} \equiv \omega \mathop{\mathsf{s}} \bar{\eta}^{}_{\mu}; \\
        & \delta B_{\mu} \equiv \omega \mathop{\mathsf{s}}  B_{\mu} =0,
    \end{align}
\end{subequations}
where $ \omega $ is an infinitesimal constant Grassmann quantity, with inverse mass dimension.  

To this end, one first verifies the important relation 
\begin{equation}\label{eq:211}
    \mathcal{L} '_{\text{gf}} + \mathcal{L}_{\text{gh}} = \mathop{\mathsf{s}} 
    \left \{ \sqrt{g} \left [ - \frac{\xi}{2} \bar{\eta}_{\mu} B^{\mu} - \bar{\eta}_{\mu} \left( h^{\mu \nu}_{; \nu} - \frac{1}{2} h_{\alpha}^{\alpha}{}^{; \mu} - \zeta \kappa \varphi \partial^{\mu} \phi    \right)\right ]\right \}  
\end{equation}
which, together  with the nilpotency property of the $ \mathsf{s} $-operator, 
\begin{equation}\label{eq:212}
    \mathop{\mathsf{s}^{2}} h^{\mu \nu} = 
    \mathop{\mathsf{s}^{2}} \varphi = 
    \mathop{\mathsf{s}^{2}} \eta^{\mu} = 
    \mathop{\mathsf{s}^{2}} \bar{\eta}^{ \mu} = 
    \mathop{\mathsf{s}^{2}} B_{\mu} =0 ,
\end{equation}
manifestly shows the BRST invariance of the total Lagrangian  Eq.~\eqref{eq:29}.

Next, one defines the effective action $ \Gamma_{\text{eff}} ( g, \phi  ) $ as 
\begin{equation}\label{eq:213}
\begin{split}
    \exp \left [ i \Gamma_{\text{eff}} ( g, \phi )\right] ={}& 
    \int \mathop{\mathcal{D} h} \mathop{\mathcal{D} \varphi}  \mathop{\mathcal{D} B} \mathop{\mathcal{D} \bar{\eta}^{}} \mathop{\mathcal{D} \eta} \\ & \times \exp i \int \mathop{d^{4} x} \left [ \mathcal{L}_{\text{tot} } ( g, \phi , h , \varphi , B , \bar{\eta}^{} , \eta ) - \mathcal{L}^{\text{inv}} ( g, \phi ) -\kappa \frac{\delta \mathcal{L}^{\text{inv}} }{\delta g_{\mu \nu} } h_{\mu \nu} - \frac{\delta \mathcal{L}^{\text{inv}} }{\delta \phi} \varphi \right ],
    \end{split}
\end{equation}
where the expression the square bracket involves terms quadratic and of higher degree in the quantum fields. To one-loop order, it is sufficient to keep only the quadratic terms.    

Similarly, one defines the background expectation value of an operator $O$ as 
\begin{equation}\label{eq:214}
    \begin{split}
    \langle O \rangle \equiv 
    & \int \mathop{\mathcal{D} h} \mathop{\mathcal{D} \varphi}  \mathop{\mathcal{D} B} \mathop{\mathcal{D} \bar{\eta}^{}} \mathop{\mathcal{D} \eta} O (g, \phi , h, \varphi , B, \bar{\eta}^{} , \eta ) \\ & \times  \exp i \int \mathop{d^{4} x} \left [ \mathcal{L}_{\text{tot} } ( g, \phi , h , \varphi , B , \bar{\eta}^{} , \eta ) - \mathcal{L}^{\text{inv}} ( g, \phi ) -\kappa \frac{\delta \mathcal{L}^{\text{inv}} }{\delta g_{\mu \nu} } h_{\mu \nu} - \frac{\delta \mathcal{L}^{\text{inv}} }{\delta \phi} \varphi \right ].
\end{split}
\end{equation}
This is invariant under BRST transformations of the fields, as these can be absorbed by a change of field variables since the Jacobian of the transformation is $1$.

Let us now examine the $ \xi $-dependence of the effective action, under an infinitesimal change $ \xi \to \xi + \Delta \xi $ (a systematic general approach is provided in \cite{Lavrov:2019nuz}). Then, using the Eqs.~\eqref{eq:28}, \eqref{eq:213} and \eqref{eq:214}, one can see that 
\begin{equation}\label{eq:215}
    \delta_{\xi} \exp i \Gamma_{\text{eff}} ( g, \phi ) = \frac{i\Delta \xi }{2} \int \mathop{d^{4} x} \sqrt{g}  \left \langle B_{\mu} B^{\mu}\right\rangle.      
\end{equation}
We next consider the operator $O = \bar{\eta}_{\mu} B^{\mu} $ which, due to Eq.~\eqref{eq:210}, satisfies the relation 
\begin{equation}\label{eq:216}
    \left \langle \mathop{\mathsf{s}} ( \bar{\eta}_{\mu} B^{\mu})\right\rangle = - \left \langle B_{\mu} B^{\mu}\right\rangle.
\end{equation}
Using the invariance of $ \left \langle \bar{\eta}_{\mu} B^{\mu}\right\rangle $  under the BRST transformations of all fields in the integrand: $ \mathop{\mathsf{s}} \left \langle \bar{\eta}_{\mu} B^{\mu}\right\rangle =0$, one gets from the Eq.~\eqref{eq:214} the Ward identity: 
\begin{equation}\label{eq:217}
    \left \langle \mathop{\mathsf{s}} ( \bar{\eta}_{\mu} B^{\mu})\right\rangle = i 
    \left \langle \bar{\eta}_{\mu} B^{\mu} \int \mathop{d^{4} x'} \left [  
   \kappa \frac{\delta \mathcal{L}^{\text{inv}} }{\delta g_{\mu \nu} } \mathop{\mathsf{s}}  h_{\mu \nu} + \frac{\delta \mathcal{L}^{\text{inv}} }{\delta \phi} \mathop{\mathsf{s}} \varphi  \right ]_{x'}\right\rangle.
\end{equation}
Combining  the Eqs.~\eqref{eq:216} and \eqref{eq:217}, one finds from Eq.~\eqref{eq:215} the equation 
\begin{equation}\label{eq:218}
    \delta_{\xi} \exp i \Gamma_{\text{eff}} (g , \phi ) = 
    \frac{ \Delta \xi }{2} \int \mathop{d^{4} x} \sqrt{g(x)}  
\left \langle (\bar{\eta}_{\mu} B^{\mu})(x) \int \mathop{d^{4} x'} \left [  
   \kappa   \frac{\delta \mathcal{L}^{\text{inv}} }{\delta g_{\mu \nu} }  \mathop{\mathsf{s}} h_{\mu \nu} + \frac{\delta \mathcal{L}^{\text{inv}} }{\delta \phi} \mathop{\mathsf{s}} \varphi  \right ]_{x'}\right\rangle.
\end{equation}

Proceeding in a similar way, one finds that under an infinitesimal change $ \zeta \to  \zeta + \Delta \zeta$:
\begin{equation}\label{eq:218a}
    \delta_{\zeta} \exp i \Gamma_{\text{eff}} (g , \phi ) = 
     - \kappa \Delta \zeta  \int \mathop{d^{4} x} \sqrt{g(x)}  
\left \langle (\bar{\eta}^{ \mu } \partial_{\mu} \phi \varphi)(x) \int \mathop{d^{4} x'} \left [  
   \kappa   \frac{\delta \mathcal{L}^{\text{inv}} }{\delta g_{\mu \nu} }  \mathop{\mathsf{s}} h_{\mu \nu} + \frac{\delta \mathcal{L}^{\text{inv}} }{\delta \phi} \mathop{\mathsf{s}} \varphi  \right ]_{x'}\right\rangle.
\end{equation}

When the background fields $ g_{\mu \nu} $ and $ \phi $ are on-shell, each functional derivative in the Eq.~\eqref{eq:218} vanishes, so that  one obtains the relation 
\begin{equation}\label{eq:219}
    \left.\delta_{\xi} \exp i \Gamma_{\text{eff}} (g, \phi ) \right|_{\text{on-shell}} = \left.\delta_{\zeta} \exp i \Gamma_{\text{eff}} (g, \phi ) \right|_{\text{on-shell}}
    = 0,
\end{equation}
which confirms the DeWitt-Kallosh theorem in the theory described by the Einstein Lagrangian Eq.~\eqref{eq:21}. In the next section, this result will be explicitly verified at one-loop order for  general values of $ \xi $ and $\zeta$.

\section{Gauge-dependence of the one-loop effective action}

On dimensional and symmetry grounds, the background counterterm Lagrangian at one loop order may be expressed in terms of the six independent structures 
\begin{equation}\label{eq:31}
    \begin{split}
    \mathcal{L}_{\text{CT}} 
    = \frac{\sqrt{g}}{16 \pi^{2} \epsilon} & \big [ c_{1} (\xi,\zeta) R^{2} + c_{2} (\xi,\zeta) R_{\mu \nu} R^{\mu \nu} + \kappa^{2} c_{3} (\xi,\zeta) ( \mathsf{D}_{\mu} \mathsf{D}^{\mu} \phi )^{2} \\ & + \kappa^{2} c_{4} (\xi,\zeta) R \partial_{\mu} \phi \partial^{\mu} \phi + \kappa^{2} c_{5} (\xi,\zeta) R^{\mu \nu} \partial_{\mu} \phi \partial_{\nu} \phi + \kappa^{4} c_{6} (\xi,\zeta)( \partial_{\mu} \phi \partial^{\mu} \phi )^{2}  \big ],
\end{split} 
\end{equation}
where $ \epsilon= (4-D)/2$,  $D$ is spacetime dimension, $c_i (\xi,\zeta)$ are dimensionless coefficients and we have taken into account the identity 
\begin{equation} \label{eq:meta31}
    \int \mathop{d^{4} x} \sqrt{g} \left(R_{\mu\nu\rho\sigma} R^{\mu\nu\rho\sigma}
       - 4 R_{\mu\nu} R^{\mu\nu}
   + R^2\right)=0. \end{equation} 
In principle, three more independent structures: $ \kappa^{2} R \phi \square \phi $, $ \kappa^{2} R^{\mu \nu} \phi \mathsf{D}_{\mu} \mathsf{D}_{\nu} \phi $, $ \kappa^{4} \phi \partial^{\mu} \phi \partial^{\nu} \phi D_{\mu} D_{\nu} \phi $ could appear in theories more general than that in Eq.~\eqref{eq:21}. 
Altogether, these nine structures (or an equivalent set), exhaust all possible independent terms expected from dimensional analysis, symmetry and the form of the interaction terms.  Such improved gravitational theories are commonly used in cosmology \cite{Barvinsky:1993zg, Kamenshchik:2014waa}.
   In our model, characterized by the gauge fixing term \eqref{eq:25}, only the six structures shown in Eq.~\eqref{eq:31} actually occur.

One could also consider more general invariants, such as 
$\kappa^{3} R \phi^{3}$ or $\kappa R_{\mu \nu} \phi\, \partial^{\mu} \phi\, \partial^{\nu} \phi$, which are odd in $\phi$. However, our simple scalar model \eqref{eq:21} is invariant under the discrete symmetry $\phi \to -\phi$. Such counterterms are therefore forbidden, as the background field method ensures that the effective action preserves the symmetries of the classical action \eqref{eq:21}. Indeed, from the Feynman rules in Appendix A, no diagram leads to an odd counterterm in $\phi$ at one-loop order. Nevertheless, spontaneous symmetry breaking can still occur in more general theories \cite{fujii:1981, *Accetta:1985du}.

\begin{figure}[ht]
    \centering
    \includegraphics[width=0.55\textwidth]{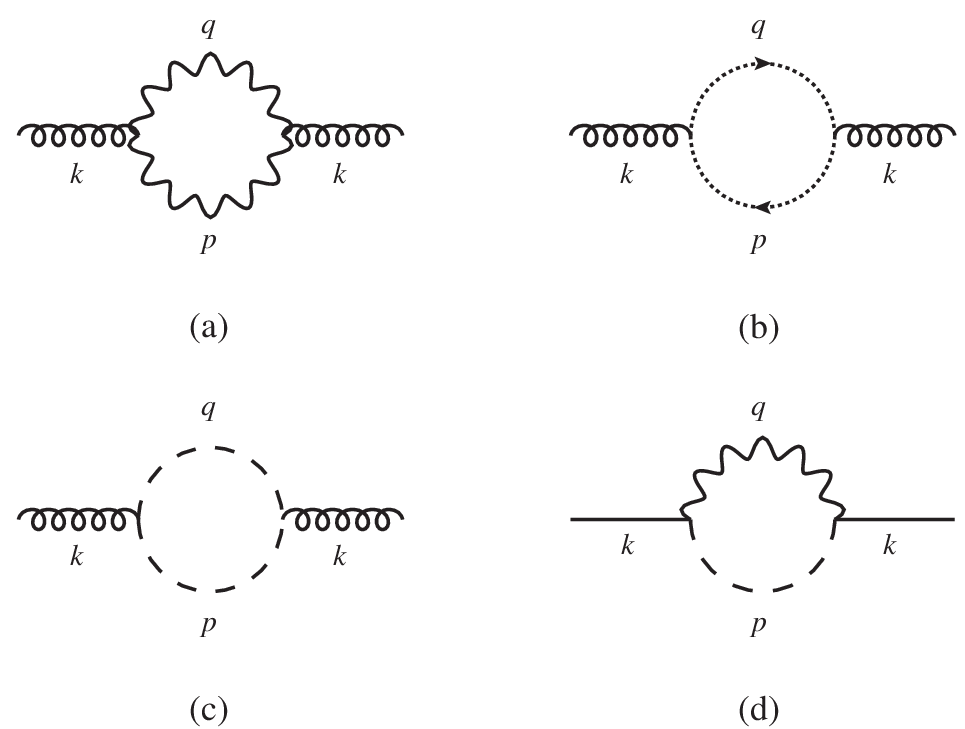}
    \caption{One-loop contributions to $ \left \langle \mathfrak{g} ^{\mu \nu} \mathfrak{g} ^{\alpha \beta}\right\rangle $ [(a), (b), (c)] and to $ \left \langle \phi \phi\right\rangle $ [(d)]. The wavy, curly, dotted, dashed and solid lines denote the quantum graviton, background graviton, ghost, quantum scalar and the background scalar, respectively. The momenta satisfies the relation: $ q =p+k$.}\label{fig:2point}
\end{figure}

The Feynman diagrams  contributing at one loop order to the background graviton self-energy $ \Pi^{\mu \nu \alpha \beta} (k) $ and the scalar self-energies $ \Pi (k)$ are shown in Fig.~\ref{fig:2point}.

As shown in the Appendix~\ref{sec:OneLoop} [see Eq.~\eqref{eq:total2}], the divergent parts of the diagrams shown in Fig.~\ref{fig:2point}(a, b, c) yield, in momentum space, the following contribution to the graviton self-energy 
\begin{equation}\label{eq:32} 
    \Pi^{\mu \nu \alpha \beta} (k) = \frac{\kappa^{2}}{16 \pi^{2} \epsilon} k^{4} \left \{ 4 c_{1} (\xi,\zeta) L^{\mu \nu} L^{\alpha \beta} + c_{2} (\xi,\zeta) \left [ L^{\mu \nu} L^{\alpha \beta} + \frac{1}{2} \left ( L^{\mu \alpha} L^{\nu \beta} + L^{\mu \beta} L^{\nu \alpha}\right )\right ] \right\}, 
\end{equation}
where $ L^{\mu \nu} = k^{\mu} k^{\nu} / k^{2} - \eta^{\mu \nu} $ and $ c_{1} (\xi,\zeta) $ and $ c_{2} (\xi,\zeta)$ are gauge-dependent coefficients given by 
\begin{equation}\label{eq:33}
    c_{1} (\xi,\zeta) = \frac{1}{80} + \frac{1}{6} ( \xi -1)^{2}, \quad c_{2} (\xi,\zeta) = \frac{43}{120} + \frac{\xi (\xi -1)}{3}.
\end{equation}
We note here that the Eq. \eqref{eq:32} is transverse to the momentum $k$, which is a consequence of the diffeomorphism invariance of  the theory.

We  can connect the above results with the counterterm Lagrangian~\eqref{eq:31} by using the relations 
\begin{align}\label{eq:34}
    R (k) ={}& \kappa k^{2} L^{\alpha \beta}  \mathfrak{g}_{\alpha \beta} + O ( \kappa^{2} ),
    \\\label{eq:35}
    R^{\mu \nu} (k) ={}& \frac{\kappa}{2} \left [ L^{\alpha \beta} k^{\mu} k^{\nu} - \frac{k^{2}}{2} \left ( L^{\mu \alpha} L^{\nu \beta} + L^{\mu \beta} L^{\nu \alpha}\right )\right ] \mathfrak{g}_{\alpha \beta} + O ( \kappa^{2} ).
\end{align}
Then, one may verify that Eq.~\eqref{eq:33} fixes the coefficients $c_1 (\xi,\zeta)$ and $c_2 (\xi,\zeta)$ which appear in the Eq.~\eqref{eq:31}. 
Likewise, the divergent part  of the scalar self-energy graph depicted in Fig.~\ref{fig:2point}(d) [see Eq.~\eqref{eq:parte2d}] fixes the coefficient $c_{3} (\xi,\zeta)$ to be 
\begin{equation}\label{eq:36}
    c_{3} (\xi,\zeta) = \frac{1}{2} + \frac{\zeta -1}{2} .
\end{equation}

\begin{figure}[ht]
    \centering
    \includegraphics[width=0.8\textwidth]{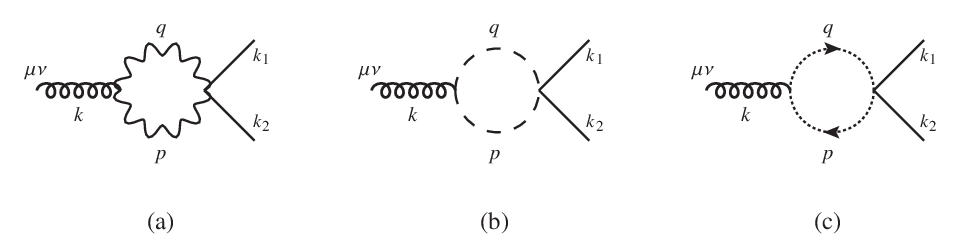}
    \includegraphics[width=0.8\textwidth]{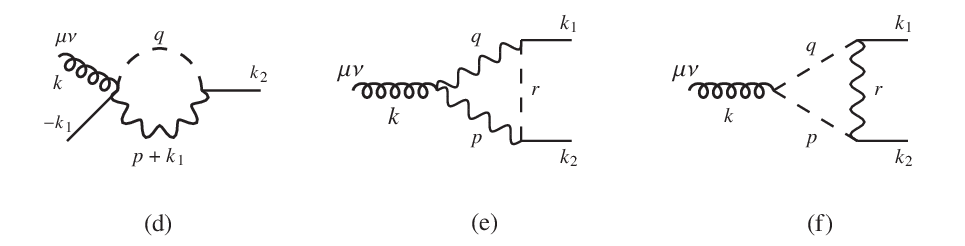}
    \caption{One-loop contributions (permutations have been omitted) to the background $3$-point function $ \langle \mathfrak{g}^{\mu \nu} (k) \phi (k_1) \phi (k_2) \rangle$. We have that $ q = p+k$,  $ r =p+k-k_1$ and $ k = k_{1} + k_{2} $.}\label{fig:3point}
\end{figure}

In order to fix the coefficients $ c_{4} (\xi,\zeta)$ and $ c_{5} (\xi,\zeta)$ in Eq.~\eqref{eq:31}, we consider the diagrams shown in Fig.~\ref{fig:3point}. 
As shown in Appendix~\ref{sec:OneLoop} [see Eq.~\eqref{eq:3pointCT}], evaluating the divergent contributions of the one-loop diagrams in Fig.~\ref{fig:3point} and using the relations given in Eqs. \eqref{eq:34} and \eqref{eq:35}, fixes the coefficients $c_4$ and $c_{5} $ as 
\begin{equation}\label{eq:37}
    c_{4} (\xi,\zeta) = - \frac{1}{12} - ( \xi -1) \frac{ \xi + 5}{12} + \frac{(\xi - \zeta ) ^2}{4} , \quad c_{5} (\xi,\zeta) = \frac{(\xi -1)( 2\xi + 5)}{6} .
\end{equation}

\begin{figure}
    \centering
\includegraphics[width=0.72\textwidth]{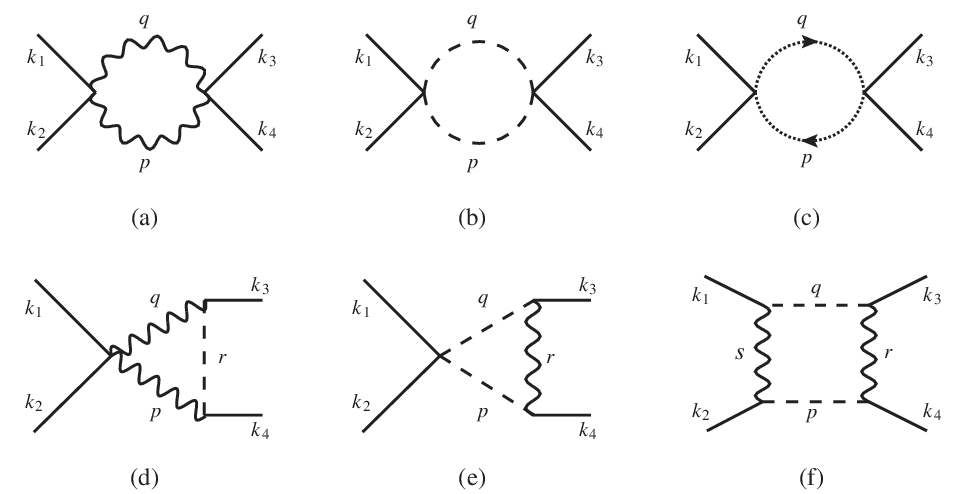} 
\caption{One-loop  contributions (permutations have been omitted) to the background $4$-point function $ \left \langle \phi (k_{1} )\phi (k_{2} )\phi (k_{3} )\phi (k_{4} )\right\rangle$. Here, $ q = p+k_{1} + k_{2} $, $ r = p + k_{4}  $ and $ s = p +k_{2} $.}\label{fig:4point}
\end{figure}

Next, we consider the divergent parts of the diagrams shown in Fig.~\ref{fig:4point}, which fix the coefficient $c_{6} (\xi,\zeta)$.
As displayed in Appendix~\ref{sec:OneLoop} [see Eq.~\eqref{eq:CT_4point}], these contributions fix the coefficient $c_{6} (\xi , \zeta )$ to be 
\begin{equation}\label{eq:38}
    c_{6}( \xi , \zeta )= \frac{1}{2} + \frac{\xi -1}{4} + \frac{(\xi - \zeta )^{2}}{8}.
\end{equation}

Substituting the coefficients~\eqref{eq:33}, \eqref{eq:36}, \eqref{eq:37} and \eqref{eq:38} in the Eq.~\eqref{eq:31}, we obtain the result
\begin{equation}\label{eq:CTLAG}
    \begin{split}
    \mathcal{L}_{\text{CT}} 
    = \frac{\sqrt{g}}{16 \pi^{2} \epsilon} & \bigg [ \frac{1}{240}\left [3 +40  ( \xi -1)^{2}\right ]  R^{2} + \frac{1}{120} \left [ 43 + 40 \xi (\xi -1)\right ]  R_{\mu \nu} R^{\mu \nu} 
                                         +  \frac{\kappa^{2}}{2} \zeta    ( \mathsf{D}_{\mu} \mathsf{D}^{\mu} \phi )^{2} 
                                        \\ & -\frac{\kappa^{2}}{12} \left [ 1 + ( \xi -1) (\xi +5) - 3(\xi - \zeta )^2\right ] R \partial_{\mu} \phi \partial^{\mu} \phi + \frac{\kappa^{2}}{6} (\xi -       1)( 2\xi +5)R^{\mu \nu} \partial_{\mu} \phi \partial_{\nu} \phi \\ & + \frac{\kappa^{4}}{8} \left [4 + 2(\xi -1) + (\xi - \zeta)^{2}\right ] ( \partial_{\mu} \phi \partial^{\mu} \phi )^{2}  \bigg ].
\end{split} 
\end{equation}
The above expression shows that the off-shell background counterterm Lagrangian \eqref{eq:CTLAG} is a gauge-dependent quantity, which reduces to that obtained in Ref. \cite{Grisaru:1975ei} for $ \xi =1$. We note here that the coefficient of $ R^{\mu \nu} \partial_{\mu} \phi \partial_{\nu} \phi $ vanishes when $ \xi =1$, which explains why this structure does not appear in the 't Hooft-Veltman counterterm Lagrangian. A comparison of our result in Eq.~\eqref{eq:CTLAG} with the one-loop counterterm found in the improved gravitational gauge theories which are used in quantum cosmology \cite{Barvinsky:1993zg, Kamenshchik:2014waa} is presented in the Appendix~\ref{sec:comparison}.

To determine the on-shell value of this quantity, we use Einstein's equations of motion 
\begin{equation}\label{eq:39}
    R_{\mu \nu} - \frac{1}{2} R g_{\mu \nu} = \frac{\kappa^{2}}{2} T_{\mu \nu}, \quad \mathsf{D}_{\mu} \mathsf{D}^{\mu} \phi =0,
\end{equation}
where the energy-momentum tensor $ T^{\mu \nu} $ corresponding to the Lagrangian \eqref{eq:21} is given by 
\begin{equation}\label{eq:310}
    T_{\mu \nu} =  \frac{1}{2} g_{\mu \nu} \partial_\alpha \phi \partial^\alpha \phi- \partial_\mu \phi \partial_\nu \phi.
\end{equation}
From the above equations, we obtain the on-shell relations 
\begin{equation}\label{eq:311}
    R_{\mu \nu} = - \frac{\kappa^{2}}{2} \partial_{\mu} \phi \partial_{\nu}  \phi, \quad R = - \frac{\kappa^{2}}{2} \partial_{\mu} \phi \partial^{\mu} \phi. 
\end{equation}
Using these relations and the Eqs.~\eqref{eq:31} and \eqref{eq:CTLAG}, we obtain for the on-shell counterterm Lagrangian the gauge-independent result (see Appendix~\ref{sec:onshell})
\begin{equation}\label{eq:312}
    \begin{split}
        \left . \mathcal{L}_{\text{CT}} \right|_{\text{on-shell}} ={}&
        \frac{\sqrt{g} \kappa^{4}}{16 \pi^{2} \epsilon} \frac{1}{4} 
        \left[c_{1} ( \xi , \zeta )+ c_{2}( \xi , \zeta ) -2 c_{4}( \xi , \zeta ) -2 c_{5} ( \xi , \zeta )+ 4 c_{6} ( \xi , \zeta )\right]
[ \partial_{\mu} \phi \partial^{\mu} \phi ]^{2}         
\\
={}&
        \frac{\sqrt{g} \kappa^{4}}{16 \pi^{2} \epsilon}\, \frac{203}{320} \ [ \partial_{\mu} \phi \partial^{\mu} \phi ]^{2} ,
\end{split}
\end{equation}
which is in agreement with the DeWitt-Kallosh theorem.

\section{Non-renormalizability of the theory}

In some gauge theories (like those with spontaneous symmetry breaking) there are non-renormalizable gauges, but there also exist gauges ('t Hooft gauges) where the renormalizability is restored. One may ask if a similar feature may also occur in the coupled scalar-gravity theory, which was proved to be, in a particular background gauge, non-renormalizable at one-loop order \cite{tHooft:1974toh, *thooft:1993}. In contrast, this theory remains non-renormalizable in a general background gauge, as may be shown in a simple way by using the DeWitt-Kallosh theorem.

    Suppose that there may exist  gauges where this theory is renormalizable in the sense that its divergences could be absorbed by field redefinitions. This property would require the relation
\begin{equation}\label{eq:41}
    \mathcal{L}^{\text{inv} } ( {g} + \Delta {g} , {\phi} +  \Delta {\phi} ) = \mathcal{L}^{\text{inv} } ( {g} , {\phi} ) + \mathcal{L}_{\text{CT}} ,
\end{equation}
where $ \mathcal{L}_{\text{CT}} $ is the counterterm Lagrangian \eqref{eq:31}. Using Eq.~\eqref{eq:21}, this relation would imply that 
\begin{equation}\label{eq:42}
\begin{split}
   & \frac{\delta \mathcal{L}^{inv} }{\delta {g}_{\mu \nu}} \Delta {g}_{\mu \nu} 
    +  
    \frac{\delta \mathcal{L}^{inv} }{\delta {\phi} } \Delta {\phi} = \\ ={}& \sqrt{g} \left \{ \left [ \frac{1}{\kappa^{2}} \left ( \frac{1}{2} g^{\mu \nu} R - R^{\mu \nu}\right ) - \frac{1}{2} \partial^{\mu} \phi \partial^{\nu} \phi + \frac{1}{4} g^{\mu \nu} \partial_{\alpha} \phi \partial^{\alpha} \phi  \right ] \Delta g_{\mu \nu} + \mathsf{D}_{\mu} \mathsf{D}^{\mu} \phi \Delta \phi \right\}
= \mathcal{L}_{\text{CT}}. 
    \end{split}
\end{equation}
On symmetry grounds and taking into account the structure of the interaction Lagrangian, one can write the variations at one-loop order in the form 
\begin{align}\label{eq:43}
    \Delta g_{\mu \nu} ={}&  \frac{\kappa^{2}}{16 \pi^{2} \epsilon} \left [ C_{1} R g_{\mu \nu} + C_{2} R_{\mu \nu} + \kappa \left ( C_{3} \partial_{\mu} \phi \partial_{\nu} \phi + C_{4} g_{\mu \nu} \partial_{\alpha} \phi \partial^{\alpha} \phi\right )\right ],\\ \label{eq:44}
    \Delta \phi ={}& \frac{\kappa^{2}}{16 \pi^{2} \epsilon} C_{5} \mathsf{D}_{\mu} \mathsf{D}^{\mu} \phi;
\end{align}
where $ C_{i} $ are some dimensionless parameters. 

Eq.~\eqref{eq:42} leads to a system of six equations expressing the five functions $ C_{i} $ in terms of the six coefficients $c_{i} $ occurring in Eq.~\eqref{eq:31}  namely 
\begin{subequations}\label{eq:45}
    \begin{align}\label{eq:45a}\allowdisplaybreaks
        2C_{1} +  C_{2} ={}&- 2c_{1} (\xi, \zeta) ,\\
        C_{2} ={}&c_{2} (\xi, \zeta), \\
        C_{5} ={}& c_{3} (\xi, \zeta),\\ 
        2 C_{1} +  C_{2} + 2 C_{3} + 4C_{4} ={}&-4c_{4} (\xi, \zeta) , \\
         C_{2} + 2C_{3} ={}&2c_{5} (\xi, \zeta), \\
        \label{eq:45f}
        C_{3} -2 C_{4} ={}&4 c_{6} (\xi, \zeta). 
    \end{align}
\end{subequations}
Solving the first five equations, yields the solution 
\begin{subequations}\label{eq:46}
    \begin{align}\label{eq:46a}
        C_{1} ={}& -  c_{1} (\xi, \zeta) + \frac{c_{2} (\xi, \zeta)}{2}, \quad C_{2} = c_{2} (\xi, \zeta) , \quad C_{3} = - \frac{c_{2} (\xi, \zeta)}{2}  + c_{5} (\xi, \zeta), \\
    C_{4} ={}& \frac{1}{4} \left[2c_{1} (\xi, \zeta) + c_{2} (\xi, \zeta)  - 4 c_{4} (\xi, \zeta) - 2 c_{5} (\xi, \zeta)\right] \quad \text{and} \quad C_{5} = c_{3} (\xi, \zeta) .
    \end{align}
\end{subequations}
Substituting  these results in  Eq. \eqref{eq:45f}, leads to the consistency condition for renormalisability 
\begin{equation}\label{eq:47}
    c_{1} (\xi, \zeta) + c_{2} (\xi, \zeta)  - 2 c_{4} (\xi, \zeta) - 2 c_{5} (\xi, \zeta) + 4 c_{6} (\xi, \zeta) =0.
\end{equation}
If this condition holds, the on-shell counter-term Lagrangian \eqref{eq:312} would vanish: 
\begin{equation}\label{eq:48}
    \mathcal{L}_{\text{CT}} |_{\text{on-shell}} = 0. 
\end{equation}
The condition \eqref{eq:47}, which implies the relation \eqref{eq:48}, is necessary for renormalizability since the counterterm Lagrangian has a different structure than that of the original Lagrangian. Furthermore, the above argument shows that the Eq.~\eqref{eq:47} is also a sufficient condition, because it follows in consequence of renormalizability.

It is worthy noting that, if Eq.~\eqref{eq:312} were gauge-dependent, one could find a gauge such that the renormalizability condition \eqref{eq:47} would be satisfied. However, according to the DeWitt-Kallosh theorem, the on-shell counterterm Lagrangian should be a gauge-independent quantity, which turns out to be nonzero.
Hence, it follows that the coupled scalar-gravity theory is not renormalizable in the above sense. 
However, as shown in Ref.~\cite{Lavrov:2019nuz}, this theory is renormalizable in the sense that all divergences can be cancelled by introducing the most general counterterms compatible with its gauge symmetry. This is achieved by using the Batalin–Vilkovisky technique \cite{Batalin:1981jr} in conjunction with the background field method.

\begin{figure}[ht]
    \includegraphics[scale=0.60]{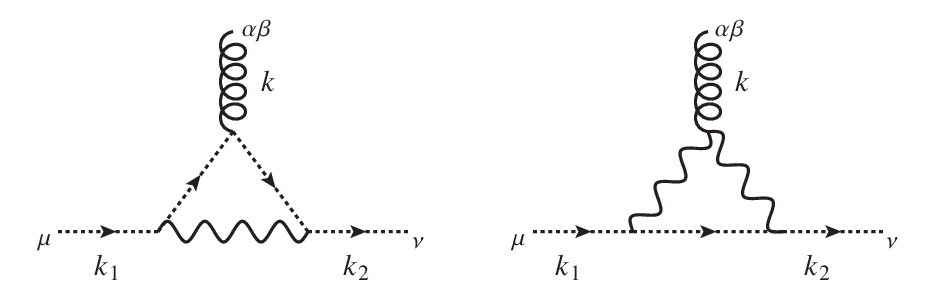}
    \caption{One-loop contributions to the ghost-background graviton-ghost vertex.}\label{fig:vertex}
\end{figure}

An interesting outcome of this non-renormalizability is the behaviour of the ghost-background graviton-ghost vertex in quantum gravity in the Landau-DeWitt gauge. This vertex is illustrated in Fig.~\ref{fig:vertex}. 

The Landau-DeWitt gauge, which is obtained in the limit $ \xi \to 0$ and $ \zeta \to 0$ of the gauge-fixing term \eqref{eq:25}, considerably  simplifies the calculations and is very useful for the understanding of many physical phenomena  \cite{Grassi:2004yq, Hamada:2015ywk}.
In Yang-Mills theory, the ghost-background gluon-ghost vertex is finite in consequence of the Ward identities in the Landau background gauge. On the other hand, as shown in Appendix~D, this no longer happens in quantum gravity, where the ghost-background graviton-ghost vertex is ultraviolet divergent in the Landau-DeWitt gauge.

\section{Discussion}

  We have examined, using the background field method, the BRST symmetry of an interacting 
quantum theory of gravity and scalar fields. This symmetry leads to an important result,known as the DeWitt-Kallosh theorem. It asserts that, on-shell, the counterterm Lagrangian ought to be
independent of the gauge-fixing parameters and of the parametrization of the fields. 
It may be useful to verify such general statements by direct calculations, which could expose some aspects that might be overlooked in formal proofs. To this end, using a general background gauge characterized by two generic gauge parameters $ \xi $ and $\zeta$, we have explicitly calculated the ultraviolet divergences at one-loop order. 

We have shown that the off-shell counterterm Lagrangian \eqref{eq:CTLAG} is a gauge-dependent quantity, which reduces to that obtained earlier by ‘t Hooft and Veltman in the particular background gauge $ \xi = \zeta=1$. Moreover, Eq.~\eqref{eq:310} is also consistent, for $ \xi = 1 $ and $ \zeta = 0$, with the results obtained more recently in \cite{Barvinsky:1993zg, Kamenshchik:2014waa}. When $ \xi =1 $, the graviton propagator is of the Feynman type, so that one can apply the same algorithm as that used in \cite{tHooft:1974toh, *thooft:1993}. On the other hand, for a general $ \xi $, this propagator is more involved, in which case one must employ a more elaborated procedure (see Appendices). We note here that in the limit $ \xi \to 0$, the contributions from some individual diagrams become singular, a feature that could pose a problem for the validity of the DeWitt-Kallosh theorem. But, when adding all terms, such singularities cancel out in consequence of the Ward identities in the Landau background gauge, which ensure that these singularities must be absent when $ \xi =0$. As a result, we obtain  a polynomial contribution which is quadratic in the gauge parameters $ \xi $ and $ \zeta $.

On the other hand, the on-shell  counterterm Lagrangian \eqref{eq:312} becomes gauge-independent, yielding a result which is in agreement with the DeWitt-Kallosh theorem. 
We applied this theorem to establish that in a general background gauge, this theory is non-renormalizable in the sense that its divergences cannot be absorbed by field redefinitions.  However, the theory is renormalizable in the broader sense that the ultraviolet divergences can be cancelled by introducing all possible counterterms allowed by its gauge symmetry.

\begin{acknowledgments}
We thank F.\ T.\ Brandt for enlightening conversations.
We thank the anonymous referee for his helpful comments and for bringing to our attention several interesting papers. 
J.\ F.\ thanks CNPq (Brazil) for financial support.  S.\ M.-F.\ thanks FAPESP for financial support. 
This study was financed, in part, by the São Paulo Research Foundation (FAPESP), Brasil. Process Number \#2025/16156-7. 
\end{acknowledgments}

\appendix

\section{Feynman rules} \label{sec:FR}
In this appendix, we show and derive some of the Feynman rules required to compute the one-loop counterterms in the Einstein-Hilbert theory coupled to a scalar field in the background field method.
Here, we follow the usual definitions and conventions used in Ref. \cite{tHooft:1974toh, *thooft:1993}: 
\begin{align}\label{eq:A0}
        R^{\mu}{}_{\nu \alpha \beta} ={}& -\partial_{\beta} \Gamma^{\mu}_{\nu \alpha} 
+ \partial_{\alpha} \Gamma^{\mu}_{\nu \beta} 
+ \Gamma^{\mu}_{\gamma \alpha} \Gamma^{\gamma}_{\nu \beta} 
- \Gamma^{\mu}_{\gamma \beta} \Gamma^{\gamma}_{\nu \alpha} \\
        R_{\mu \nu \alpha \beta} ={}& g_{\mu \gamma} R^{\gamma}{}_{\nu \alpha \beta}, 
\quad 
R_{\nu \beta} = R^{\mu}{}_{\nu \beta \mu}, 
\quad 
R = R_{\nu \beta} g^{\nu \beta}.
\end{align}
Moreover, we will use that 
\begin{equation}\label{eq:A1}
    g_{\mu \nu} = \eta_{\mu \nu} + \kappa \mathfrak{g}_{\mu \nu},
\end{equation}
where $ \eta_{\mu \nu} $ is a flat background and $ \mathfrak{g}_{\mu \nu} $ is the background graviton field. 

In order to derive the Feynman rules, we use the definitions \eqref{eq:22} and \eqref{eq:A1} 
\begin{subequations}\label{eq:A2}
    \begin{align}\label{eq:A2a}
        \bar{g}^{\mu \nu} ={}& \eta^{\mu \nu} - \kappa \mathfrak{g}^{\mu \nu}   
        - \kappa \eta^{\mu \alpha} \eta^{\nu \beta} h_{\alpha \beta} 
        + \kappa^{2} \eta^{\mu \alpha} \mathfrak{g}^{\nu \beta} h_{\alpha \beta} 
        + \kappa^{2} \mathfrak{g}^{\mu \alpha} \eta^{\nu \beta} h_{\alpha \beta} 
        + \kappa^{2} h^{\mu}_{\alpha} h^{\alpha \nu}  
        , \\
        \sqrt{\bar{g}} ={}& 
        1 + \frac{\kappa}{2}  \mathfrak{g}_{\mu}^{\mu} 
        + \frac{\kappa }{2} \eta^{\mu \nu} h_{\mu \nu}
        - \frac{\kappa^{2} }{2} \mathfrak{g}^{\mu \nu} h_{\mu \nu}          + \frac{ \kappa^{2} }{4} 
        \mathfrak{g}_{\mu}^{\mu} \eta^{\alpha \beta} h_{\alpha \beta}
        + \frac{\kappa^{2}}{8} \left ( h_{\alpha}^{\alpha}{}^{2} -2h^{\alpha}_{\beta} h^{\beta}_{\alpha} \right ), \\
        \tensor{\Gamma}{_{\mu \nu}^{\lambda}} ={}& 
        \frac{\kappa}{2} \eta^{\lambda \rho} ( \partial_{\nu} \mathfrak{g}_{\rho \mu} + \partial_{\mu} \mathfrak{g}_{\rho \nu} - \partial_{\rho} \mathfrak{g}_{\mu \nu} )
        ,\\
        \mathsf{D}_{\sigma} h_{\mu \nu} ={}& 
 \partial_\sigma h_{\mu\nu} 
-
        \frac{\kappa}{2} \eta^{\alpha \rho}
        \left[ ( \partial_{\sigma} \mathfrak{g}_{\rho \mu} + \partial_{\mu} \mathfrak{g}_{\rho \sigma} - \partial_{\rho} \mathfrak{g}_{\mu \sigma} )
            h_{\alpha\nu}
+ 
         ( \partial_{\sigma} \mathfrak{g}_{\rho \nu} + \partial_{\nu} \mathfrak{g}_{\rho \sigma} - \partial_{\rho} \mathfrak{g}_{\nu \sigma} )
        h_{\mu\alpha}\right],
    \end{align}
\end{subequations}
in the effective EH action 
\begin{equation}\label{eq:effact}
    i S = i \int \mathop{d^{4} x} \left ( \mathcal{L}^{\text{inv} } + \mathcal{L}_{\text{gf}} + \mathcal{L}_{\text{gh}}\right ),
\end{equation}
where $ \mathcal{L}^{\text{inv} } $, $ \mathcal{L}_{\text{gf}} $, $ \mathcal{L}_{\text{gh}} $ are respectively the invariant \eqref{eq:21}, gauge-fixing \eqref{eq:25} and ghost \eqref{eq:27} Lagrangians.

Consider the perturbation $g_{\mu\nu} \rightarrow \tilde{g}_{\mu \nu} =\eta_{\mu\nu} + \kappa \tilde{\mathfrak{g}}_{\mu\nu}$. One can show that, in momentum space, 
\begin{subequations}
\begin{equation}\label{vR}
\sqrt{\tilde{ g}}\tilde{ R} )
= \kappa q^2 L^{\alpha \beta } \, \tilde{\mathfrak{g}}_{\alpha\beta} + {\cal O}(\kappa^2),
\end{equation}
\begin{equation}\label{vRmn}
    \sqrt{\tilde{ g}} \,\tilde{ R}^{\mu\nu})
= 
\frac{\kappa}{2} \left[
L^{\rho\sigma}(q)\,{{q}^{\mu }{q}^{\nu }}
-\frac{q^2}{2} [L^{\mu \sigma}(q) L^{\rho\nu}(q) + L^{\mu \rho}(q) L^{\sigma\nu}(q) ]
\right] \tilde{\mathfrak{g}}_{\rho\sigma}(q) + {\cal O}(\kappa^2).
\end{equation}
\end{subequations}
where
\begin{equation}\label{transvL}
L^{\mu\nu}(q) = \frac{q^\mu q^\nu}{q^2} -  \eta^{\mu\nu}.
\end{equation}
We also have that:
\begin{subequations}\label{Rinvariants}
\begin{equation}\label{RinvA}
\sqrt{\tilde{ g}}\tilde{ R}^2
= \kappa^2 q^4 L^{\alpha \beta }(q) L^{\mu \nu }(q)
\tilde{\mathfrak{g}}_{\mu\nu} \tilde{\mathfrak{g}}_{\alpha\beta}
+ {\cal O}(\kappa^3),
\end{equation}
\begin{equation}\label{RinvB}
\sqrt{\tilde{ g}}\tilde{ R}_{\mu\nu}\tilde{ R}^{\mu\nu})
= \kappa^2 q^4\biggl[
\frac{1}{4} L^{\alpha \beta }(q) L^{\mu \nu}(q)
+\frac{1}{8} L^{\nu \alpha }(q) L^{\mu \beta}(q)
+\frac{1}{8} L^{\mu \alpha }(q) L^{\nu \beta }(q)
\biggr]
\tilde{\mathfrak{g}}_{\mu\nu} \tilde{\mathfrak{g}}_{\alpha\beta} + {\cal O}(\kappa^3).
\end{equation}
\end{subequations}

\subsection{Propagators for the quantum fields: $h_{\mu\nu}$ and $ \bar{\eta}^{}_{\mu} $, $\eta_{\mu} $}

The quadratic form in the quantum graviton field is given by
\begin{equation}
\int d^4 x h_{\mu\nu} B^{\mu\nu \, \alpha\beta} h_{\alpha\beta},
\end{equation}
where 
\begin{align}\label{quadratic}
    B^{\mu\nu\,\alpha\beta} ={}
    - \bigg[& \frac{1}{\xi}\left({\frac{1}{2} \partial^{\mu } \partial^{\nu } \eta^{\alpha
   \beta }-\partial^{\beta } \partial^{\nu } \eta^{\alpha \mu
   }+\frac{1}{2} \partial^{\alpha } \partial^{\beta } \eta^{\mu \nu
   }-\frac{1}{4} \partial^2 \eta^{\alpha \beta } \eta^{\mu \nu }}\right)
\\ & +\partial^{\beta } \partial^{\nu } \eta^{\alpha \mu
   }-\partial^{\alpha } \partial^{\beta } \eta^{\mu \nu }-\frac{1}{2}
   \partial^2 \eta^{\alpha \mu } \eta^{\beta \nu }+\frac{1}{2} \partial^2 \eta^{\alpha \beta }
   \eta^{\mu \nu }\bigg]
\nonumber \\
&+ \mbox{symmetrizations } \mu\leftrightarrow\nu \mbox{ and } \alpha\leftrightarrow\beta
\nonumber \\
&+ \mbox{permutation } (\mu,\nu) \leftrightarrow (\alpha,\beta).
\end{align}

Going to momentum space ($\partial\rightarrow i p$) and computing $i[B^{\mu\nu\alpha\beta}(p)]^{-1}$ with computer algebra (throughout this work we have used Package-{\sf X} \cite{Patel:2015tea}), we obtain
\begin{align}\label{propquantumh}
    {\cal D}_{\mu\nu\,\alpha\beta}(p) ={}-\frac{i}{p^2 + i \epsilon }  \Biggl[
   \eta_{\alpha \nu } \eta_{\beta \mu }
+  \eta_{\alpha \mu } \eta_{\beta \nu }
- \eta_{\alpha \beta } \eta_{\mu \nu}
+&\frac{\xi -1}{ { p^2}} \biggl(
p_{\beta } p_{\nu } \eta_{\alpha
    \mu }
+ p_{\beta } p_{\mu } \eta_{\alpha
    \nu }
+ p_{\alpha } p_{\nu } \eta_{\beta
    \mu }
+ p_{\alpha } p_{\mu } \eta_{\beta
    \nu }\biggr)\Biggl].
\end{align}

Similarly, from Eq. \eqref{eq:27} we obtain the following expression for the ghost propagator
\begin{equation}
{\cal D}_{\mu\nu}(p) = -i \frac{\eta^{\mu\nu}}{p^2+i\epsilon}.
\end{equation}

\subsection{Vertex $\mathfrak g h h$}

The interactions terms $ \mathfrak{g} h h $ arises from \cite{Brandt:2022und}
\begin{equation}\label{eq:A14A}
    \mathcal{L}^{(2)}  = -\frac{1}{2} T^{\alpha \mu \nu \beta \gamma \delta} \sqrt{g} \mathsf{D}_{\alpha} {h}_{\mu \nu} \mathsf{D}_{\beta} {h}_{\gamma \delta} - \frac{\sqrt{g}}{2} \left [ R \left ( \frac{1}{4} h^{2} - \frac{1}{2} h_{\mu \nu} h^{\mu \nu}\right ) + R^{\mu \nu} \left ( 2 h^{\alpha}_{\mu} h_{\nu \alpha} - h h_{\mu \nu}\right )\right ] ,
\end{equation}
where $ h \equiv  h_{\mu \nu} \mathfrak{h}^{\mu \nu} $ and
\begin{align}
    T^{\alpha\mu\nu\,\beta\gamma\delta} ={}&
     {H^{(0)}}^{\alpha\mu\nu\beta\gamma\delta} 
+
 \frac{1}{2} {H^{(1)}}^{\alpha\mu\nu\beta\gamma\delta}
-\frac{1}{2} {H^{(1)}}^{\alpha\mu\gamma\beta\nu\delta}
+{H^{(1)}}^{\alpha\mu\beta\gamma\nu\delta}
-{H^{(1)}}^{\alpha\mu\nu\delta\gamma\beta}
\nonumber \\ &+
\frac{1}{\xi}\left(
 {H^{(1)}}^{\alpha\mu\beta\nu\gamma\delta}
-\frac{1}{2} {H^{(1)}}^{\alpha\mu\gamma\nu\beta\delta}
-\frac{1}{2} {H^{(1)}}^{\alpha\mu\beta\gamma\nu\delta}
+\frac{1}{4} {H^{(1)}}^{\alpha\mu\gamma\beta\nu\delta}
\right)
 + {\cal O}(\kappa^2),
\end{align}
with
\begin{eqnarray}\label{eq:H0}
{H^{(0)}}^{\alpha\mu\nu\beta\gamma\delta}&\equiv&
\frac{1}{2} \eta^{\alpha\beta}\eta^{\mu\gamma} \eta^{\nu\delta}
-\frac{1}{2} \eta^{\alpha\beta}\eta^{\mu\nu} \eta^{\gamma\delta}
+\eta^{\alpha\gamma}\eta^{\mu\nu} \eta^{\beta\delta}
-\eta^{\alpha\delta}\eta^{\mu\gamma} \eta^{\nu\beta}
\nonumber \\ &+&
\frac{1}{\xi}\left(
 \eta^{\alpha\nu}\eta^{\mu\gamma} \eta^{\beta\delta}
-\frac{1}{2} \eta^{\alpha\nu}\eta^{\mu\beta} \eta^{\gamma\delta}
-\frac{1}{2} \eta^{\alpha\gamma}\eta^{\mu\nu} \eta^{\beta\delta}
+\frac{1}{4}\eta^{\alpha\beta}\eta^{\mu\nu} \eta^{\gamma\delta}
\right)
\end{eqnarray}
and
\begin{eqnarray}
{H^{(1)}}^{\alpha\mu\nu\beta\gamma\delta}&=&
\kappa\left(-\delta^{\alpha}_{\mu_1}\delta^{\beta}_{\nu_1}  \eta^{\mu\gamma}  \eta^{\nu\delta} 
-\eta^{\alpha\beta}  \delta^{\mu}_{\mu_1}\delta^{\gamma}_{\nu_1}  \eta^{\nu\delta} 
-\eta^{\alpha\beta}  \eta^{\mu\gamma}  \delta^{\nu}_{\mu_1}\delta^{\delta}_{\nu_1}\right)\mathfrak g^{\mu_1\nu_1}.
\nonumber \\ &\equiv &
\kappa\; G^{\mu_1\nu_1\,\alpha\mu\nu\beta\gamma\delta}
\mathfrak g_{\mu_1\nu_1}.
\end{eqnarray}

The total result for the $\mathfrak g_{\mu_1\nu_1} h_{\mu_2\nu_2} h_{\mu_3\nu_3}$ interactions vertex is given by
\begin{equation}\label{eq:vertexhhh}
    \vcenter{\hbox{\includegraphics[scale=0.6]{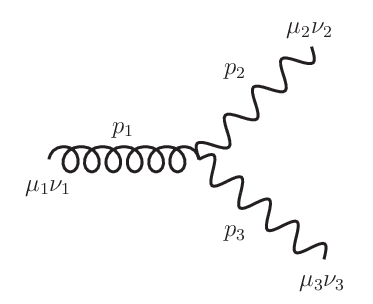}}}
    \quad 
    \begin{aligned}     (-1)i\frac{\kappa}{2}
& \Bigg\{
  \delta^{\mu_2}_{\mu} \delta^{\nu_2}_{\nu} \delta^{\mu_3}_{\gamma} \delta^{\nu_3}_{\delta} {p_2}{}_\alpha {p_3}{}_\beta  \bigg[
     \frac{1}{2} \eta^{\mu_1\nu_1} {H^{(0)}}^{\alpha\mu\nu\beta\gamma\delta}
   +    \frac{1}{2} G^{\mu_1\nu_1\,\alpha\mu\nu\beta\gamma\delta}
\\ & -\frac{1}{2} G^{\mu_1\nu_1\,\alpha\mu\gamma\beta\nu\delta}
+G^{\mu_1\nu_1\,\alpha\mu\beta\gamma\nu\delta}
-G^{\mu_1\nu_1\,\alpha\mu\nu\delta\gamma\beta}
\\ & +
\frac{1}{\xi}\left(
 G^{\mu_1\nu_1\,\alpha\mu\beta\nu\gamma\delta}
-\frac{1}{2} G^{\mu_1\nu_1\,\alpha\mu\gamma\nu\beta\delta}
-\frac{1}{2} G^{\mu_1\nu_1\,\alpha\mu\beta\gamma\nu\delta}
+\frac{1}{4} G^{\mu_1\nu_1\,\alpha\mu\gamma\beta\nu\delta}
\right)
\bigg]
\nonumber \\
&  
+{H^{(0)}}^{\alpha\mu\nu\beta\gamma\delta}
\Biggr[\Biggl(\frac{1}{2}
  ( \delta_{\mu}^{ \mu_3} \delta_{\nu}^{\nu_3} {\bar{p}_1}{}_{\gamma }
   {p_3}{}_{\alpha } \delta_{\beta}^{ \nu_1} \delta_{\delta}^{ \mu_2} \eta^{\mu_1\nu_2}
   + \delta_{\mu}^{\mu_3} \eta^{\nu_1\nu_2} \delta_{\nu}^{\nu_3} {\bar{p}_1}{}_{\beta } {p_3}{}_{\alpha }
   \delta_{\gamma}^{ \mu_1} \delta_{\delta}^{ \mu_2}
\\ & - \delta_{\mu}^{\mu_3} \delta_{\nu}^{\nu_3}
   {\bar{p}_1}^{\nu_2} {p_3}{}_{\alpha } \delta_{\beta}^{ \mu_1} \delta_{\gamma}^{ \mu_2} \delta_{\delta}^{ \nu_1}
   )
   +
    \gamma \leftrightarrow \delta\Biggr)
+ (\alpha,\mu,\nu) \leftrightarrow(\beta,\gamma,\delta)
\Biggl]
   \\
   & + \bar{p}_1^2 \, L^{\mu_1\nu_1}(\bar{p}_1)
\left(
 \frac{1}{4} \eta^{\mu_2\nu_2} \eta^{\mu_3\nu_3} - \frac{1}{2}\eta^{\mu_2\mu_3}\eta^{\nu_2\nu_3}\right)
 \nonumber \\ &+
 \frac{1}{2}\left[
L^{\mu_1\nu_1}(\bar{p}_1) {\bar{p}_1}^\mu {\bar{p}_1}^\nu - \frac{\bar{p}_1^2}{2}\left(
L^{\mu\mu_1}(\bar{p}_1) L^{\nu\nu_1}(\bar{p}_1) + L^{\nu\mu_1}(\bar{p}_1) L^{\mu\nu_1}(\bar{p}_1)
\right)
\right]
\nonumber\\ & 
 \left(
2 \eta^{\nu_3\mu_2} \delta_{\mu}^{\nu_2} \delta_{\nu}^{\mu_3}-\eta^{\mu_2\nu_2}\delta_\mu^{\mu_3}\delta_\nu^{\nu_3}
\right)
\bigg\}
+ \mbox{ symmetrizations } \mu_i\leftrightarrow\nu_i
\nonumber \\
&+ \mbox{ permutation } (\mu_2,\nu_2,p_2) \leftrightarrow (\mu_3,\nu_3,p_3).
\end{aligned}
\end{equation}

\subsection{The vertex $\bar{\eta}  \mathfrak{g}  \eta$}
The interaction terms coming from ghost Lagrangian given in Eq. \eqref{eq:27} reads 
\begin{align}\label{eq:LGHOSt}
 &  \kappa \Bigg\{\frac{1}{2}  \mathfrak{g}^\lambda_\lambda \bar{\eta}_{\mu} \partial^2 \eta^\mu
-\bar{\eta}_{\mu} \left( \mathfrak{g}^{\beta\lambda}\eta^{\mu\nu} +  \mathfrak{g}^{\mu\nu}\eta^{\beta\lambda}\right) \partial_\beta\partial_\lambda \eta_\nu
\nonumber \\
&+\frac{1}{2} \eta^{\mu\nu}\eta^{\beta\lambda}
\bigg[
(\partial_\beta \bar{\eta}_\mu) \eta^{\rho\sigma}
\left(
\partial_\lambda \mathfrak{g}_{\sigma\nu}+\partial_\nu \mathfrak{g}_{\sigma\lambda}-\partial_\sigma \mathfrak{g}_{\nu\lambda}\right)\eta_{\rho}
\nonumber \\
& \qquad \qquad  \quad -\bar{\eta}_\mu\eta^{\rho\sigma}
\left(
\partial_\beta \mathfrak{g}_{\sigma\lambda}+\partial_\lambda \mathfrak{g}_{\sigma\beta}-\partial_\sigma \mathfrak{g}_{\lambda\beta}\right)\partial_\rho \eta_{\nu}
\nonumber \\
& \qquad \qquad \quad  -\bar{\eta}_\mu\eta^{\rho\sigma}
\left(
\partial_\beta \mathfrak{g}_{\sigma\nu}+\partial_\nu \mathfrak{g}_{\sigma\beta}-\partial_\sigma \mathfrak{g}_{\nu\beta}\right)\partial_\lambda \eta_{\rho}
\bigg]\Bigg\} - \bar{\eta}_\mu R^{\mu\nu} \eta_\nu .
\end{align}
From \eqref{eq:LGHOSt}, we obtain the following momentum space interaction vertex $ h_{\alpha\beta} \bar{\eta}_\mu \eta_\nu$:
\begin{equation}\label{eq:vertexghost}
    \vcenter{\hbox{\includegraphics[scale=0.6]{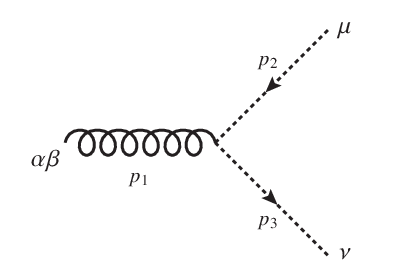}}}
    \qquad
    \begin{aligned}
    & i\kappa\bigg\{
-\frac{1}{2} p_3^2 \eta^{\alpha\beta} \eta^{\mu\nu} 
+p_3^{\alpha}p_3^{\beta}\eta^{\mu\nu} + p_3^2 \eta^{\mu\alpha} \eta^{\nu\beta} 
\nonumber \\ &+\frac{1}{2}\left(
\bar{p}_1^{\nu} p_2^{\beta} \eta^{\mu\alpha}-\bar{p}_1^{\mu} p_2^{\beta} \eta^{\nu\alpha}
-\bar{p}_1\cdot p_3 \eta^{\mu\beta} \eta^{\nu\alpha}
\right)
\nonumber \\ &+ \frac{1}{2}\left(
2 \, p_3^{\alpha} \bar{p}_1^{\beta} \eta^{\mu\nu}-\bar{p}_1\cdot p_3 \eta^{\mu\nu}\eta^{\alpha\beta}
\right)
\nonumber \\ &-\frac{1}{2}\left(
\bar{p}_1^{\nu} p_2^{\beta} \eta^{\mu\alpha}-\bar{p}_1^{\mu} p_2^{\alpha} \eta^{\nu\beta}
-\bar{p}_1\cdot p_3 \eta^{\mu\beta} \eta^{\nu\alpha}
\right)
\nonumber \\
&-\frac{1}{2}\left[
\bar{p}_1^{\alpha}\left(
\bar{p}_1^{\mu}\eta^{\nu\beta}+\bar{p}_1^{\nu}\eta^{\mu\beta}\right) - \bar{p}_1^{\mu} \bar{p}_1^{\nu} \eta^{\alpha\beta} - \bar{p}_1^2\eta^{\mu\alpha}\eta^{\nu\beta}
\right]
\bigg\}
\nonumber \\
&+ \mbox{ symmetrizations } \alpha\leftrightarrow\beta ,
\end{aligned}
\end{equation}
where we have used Eq. \eqref{vRmn}.

\subsection{Gravity coupled to a scalar field}
Now, we will consider the coupling between the graviton and a scalar field. From the classical Lagrangian
\begin{equation}\label{eq:scalarcoupled}
    - \sqrt{\bar{g} } \frac{\bar{g}^{\mu \nu}}{2} \partial_{\mu} \bar{\phi} \partial_{\nu} \bar{\phi},
\end{equation}
we obtain the scalar propagator in the momentum space:
\begin{equation}\label{eq:propscalar}
    -i \frac{1}{k^{2}} .
\end{equation}
The interactions terms arise from the Lagrangian \eqref{eq:213}:
\begin{equation}\label{eq:scalarcoupled2}
    - \sqrt{\bar{g} } \frac{\bar{g}^{\mu \nu}}{2} \partial_{\mu} \bar{\phi} \partial_{\nu} \bar{\phi}
    + \kappa \frac{ \zeta }{ \xi} \sqrt{g} \left( h^{\mu \nu}_{; \nu} - \frac{1}{2} h_{\alpha}^{\alpha}{}^{; \mu} \right)   \varphi \partial_{\mu} \phi     
    - \kappa^{2}\frac{ \zeta^{2} }{2 \xi} \sqrt{g}  
\left(\varphi \partial^{\mu} \phi    \right)^{2}
      - \zeta \kappa^{2}\sqrt{g} \bar{\eta}^{ \mu}   \partial_{\mu} \phi \partial_{\nu} \phi \eta^{\nu}. 
\end{equation}

Consider the interaction terms $ \mathfrak{g} h \phi \varphi $. Arising from the first term of Eq.~\eqref{eq:scalarcoupled2}, we have
\begin{equation}\label{eq:4vertexa}
    \begin{split}
        - \frac{\kappa^{2}}{4}  \big[& 4\eta^{\mu \alpha} \mathfrak{g}^{\nu \beta} h_{\alpha \beta} 
   +  4\mathfrak{g}^{\mu \alpha} \eta^{\nu \beta} h_{\alpha \beta} 
   + \mathfrak{g}_{\pi}^{\pi} \eta^{\alpha \beta} \eta^{\mu \nu} h_{\alpha \beta}
     \\ & - 2 (\mathfrak{g}^{\alpha \beta } h_{\alpha \beta } \eta^{\mu \nu} 
   + \eta^{\mu \alpha} \eta^{\nu \beta} h_{\alpha \beta} \mathfrak{g}_{\pi}^{\pi} +  \mathfrak{g}^{\mu \nu } \eta^{\alpha \beta} h_{\alpha \beta}) 
   \big] \partial_{\mu} \phi \partial_{\nu} \varphi.
    \end{split}
\end{equation}
The remaining terms comes from the second term of Eq.~\eqref{eq:scalarcoupled2}, which can be written as 
\begin{equation}\label{eq:term2}
    \kappa \frac{ \zeta }{ \xi} \sqrt{g} T^{\mu \nu \, \alpha \beta}  h_{\alpha \beta ; \nu}   \varphi \partial_{\mu} \phi
    \equiv 
    \kappa \frac{ \zeta }{ \xi} \sqrt{g} ( g^{\alpha \mu} g^{\beta \nu} - \frac{1}{2} g^{\mu \nu} g^{\alpha \beta} ) h_{\alpha \beta ; \nu}   \varphi \partial_{\mu} \phi. 
\end{equation}
Using Eq.~\eqref{eq:A2}, we obtain from the determinant the contribution
\begin{equation}\label{eq:4vertexdet}
    + \frac{\kappa^{2}}{2} \frac{\zeta}{\xi} \mathfrak{g}_{\beta}^{\beta} \left( \partial_{\nu} h^{\dot\mu \dot\nu} - \frac{1}{2} \partial^{\dot\mu} h_{\alpha \beta } \eta^{\alpha \beta}  \right)   \varphi \partial_{\mu} \phi,  
\end{equation}
coming from the tensor $T^{\mu \nu \, \alpha \beta} $, we obtain 
\begin{equation}\label{eq:4vertexT}
- \kappa^{2} \frac{\zeta}{\xi} 
   \left [ \eta^{\mu \alpha} \mathfrak{g}^{\lambda \beta} + \mathfrak{g}^{\mu \alpha} \eta^{\lambda \beta} - \frac{1}{2} (\eta^{\mu \lambda} \mathfrak{g}^{\alpha \beta} + \mathfrak{g}^{\mu \lambda} \eta^{\alpha \beta})\right ] 
   \partial_{\lambda} h_{\alpha \beta} 
\varphi \partial_{\mu} \phi
\end{equation}
and, finally, from the covariant derivative: 
\begin{equation}\label{eq:4vertexb}
   - \frac{\kappa^{2}}{2} \frac{\zeta}{\xi} 
   \left [ \eta^{\mu \alpha} \eta^{\lambda \beta}  - \frac{1}{2} \eta^{\mu \lambda} \eta^{\alpha \beta} \right] 
       \eta^{\sigma \rho}
        \left[ ( \partial_{\lambda} \mathfrak{g}_{\rho \alpha} + \partial_{\alpha} \mathfrak{g}_{\rho \lambda} - \partial_{\rho} \mathfrak{g}_{\alpha \lambda} )
            h_{\sigma \beta}
+ 
         ( \partial_{\lambda} \mathfrak{g}_{\rho \beta} + \partial_{\beta} \mathfrak{g}_{\rho \lambda} - \partial_{\rho} \mathfrak{g}_{\beta \lambda} )
        h_{\alpha\sigma}\right] 
\varphi \partial_{\mu} \phi.
\end{equation}
Note that, we have introduced the notation $ T^{\dot{\mu}_{1} \cdots \dot \mu_{n} } \equiv \eta^{\mu_{1} \nu_{2} } \cdots \eta^{\mu_{n} \nu_{n} }   T_{\nu_{1} \nu_{2} \cdots \nu_{n} } $. 

Adding all contributions, the vertex $ \mathfrak{g}_{\alpha \beta} (p_{4} )h_{\mu \nu} ( p_{3} )\varphi ( p_{1} )\phi (p_{2} )$ follows in momenta space as
\begin{equation}\label{eq:Vbhhpbp}
    \vcenter{\hbox{\includegraphics[scale=0.6]{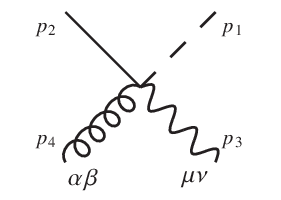}}}
    \begin{aligned}
        -i\frac{\kappa^2}{4 \xi }& \big[-\xi  p_{1}{}^{\beta} p_{2}{}^{\nu } \eta^{\alpha\mu }-\xi  p_{1}{}^{\beta} p_{2}{}^{\mu } \eta^{\alpha\nu }-\xi  p_{1}{}^{\alpha} p_{2}{}^{\nu } \eta^{\beta\mu }+\xi  p_{1}{}^{\nu } \left(p_{2}{}^{\mu } \eta^{\alpha\beta}-p_{2}{}^{\beta} \eta^{\alpha\mu }-p_{2}{}^{\alpha} \eta^{\beta\mu }\right) 
                              \\ & -\xi  p_{1}{}^{\alpha} p_{2}{}^{\mu } \eta^{\beta\nu }+\xi  p_{1}{}^{\mu } \left(p_{2}{}^{\nu } \eta^{\alpha\beta}-p_{2}{}^{\beta} \eta^{\alpha\nu }-p_{2}{}^{\alpha} \eta^{\beta\nu }\right)+\xi  p_{2}{}^{\alpha} p_{1}{}^{\beta} \eta^{\mu \nu }+\xi  p_{1}{}^{\alpha} p_{2}{}^{\beta} \eta^{\mu \nu } 
                              \\ & -\xi  p_{1}{} \cdot p_{2}{} \eta^{\alpha\beta} \eta^{\mu \nu }+\xi  p_{1}{} \cdot p_{2}{} \eta^{\alpha\nu } \eta^{\beta\mu }+\xi  p_{1}{} \cdot p_{2}{} \eta^{\alpha\mu } \eta^{\beta\nu }+\zeta  p_{3}{}^{\mu } p_{2}{}^{\nu } \eta^{\alpha\beta}+\zeta  p_{2}{}^{\mu } p_{3}{}^{\nu } \eta^{\alpha\beta} \\ & -\zeta  p_{3}{}^{\beta} p_{2}{}^{\nu } \eta^{\alpha\mu }-\zeta  p_{2}{}^{\beta} p_{3}{}^{\nu } \eta^{\alpha\mu }-\zeta  p_{3}{}^{\beta} p_{2}{}^{\mu } \eta^{\alpha\nu }-\zeta  p_{2}{}^{\beta} p_{3}{}^{\mu } \eta^{\alpha\nu }-\zeta  p_{3}{}^{\alpha} p_{2}{}^{\nu } \eta^{\beta\mu } \\ & -\zeta  p_{2}{}^{\alpha} p_{3}{}^{\nu } \eta^{\beta\mu }-\zeta  p_{3}{}^{\alpha} p_{2}{}^{\mu } \eta^{\beta\nu }-\zeta  p_{2}{}^{\alpha} p_{3}{}^{\mu } \eta^{\beta\nu }+\zeta  p_{3}{}^{\alpha} p_{2}{}^{\beta} \eta^{\mu \nu }+\zeta  p_{2}{}^{\alpha} p_{3}{}^{\beta} \eta^{\mu \nu } \\ & +\zeta  p_{4}{}^{\mu } p_{2}{}^{\nu } \eta^{\alpha\beta}+\zeta  p_{2}{}^{\mu } p_{4}{}^{\nu } \eta^{\alpha\beta}-\zeta  p_{4}{}^{\beta} p_{2}{}^{\nu } \eta^{\alpha\mu }-\zeta  p_{4}{}^{\beta} p_{2}{}^{\mu } \eta^{\alpha\nu }-\zeta  p_{4}{}^{\alpha} p_{2}{}^{\nu } \eta^{\beta\mu } \\ & -\zeta  p_{4}{}^{\alpha} p_{2}{}^{\mu } \eta^{\beta\nu }-\zeta  p_{2}{} \cdot p_{3}{} \eta^{\alpha\beta} \eta^{\mu \nu }+\zeta  p_{2}{} \cdot p_{3}{} \eta^{\alpha\nu } \eta^{\beta\mu }+\zeta  p_{2}{} \cdot p_{3}{} \eta^{\alpha\mu } \eta^{\beta\nu }\big].
    \end{aligned}
\end{equation}

From the first term of Eq.~\eqref{eq:scalarcoupled2}, we obtain the interaction terms $ \mathfrak{g} \varphi \varphi $, which leads to the vertex $ \mathfrak{g}_{\mu \nu} \varphi( p_{1} )\varphi ( p_{2} )$:
\begin{equation}\label{eq:Vhpp}
    \vcenter{\hbox{\includegraphics[scale=0.6]{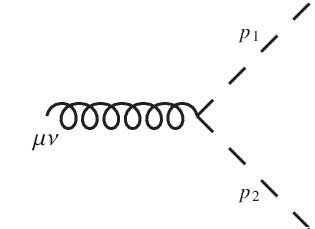}}}
    \qquad 
    i \frac{\kappa}{2}\left ( \eta_{\mu \nu} {p}_{1} \cdot p_{2}   - {p}_{1 \mu} {p}_{2 \nu} - {p}_{1 \nu} {p}_{2 \mu}\right ).
\end{equation}

This following interaction terms from Eq.~\eqref{eq:scalarcoupled2}
\begin{equation}\label{eq:lagbpph}
    \kappa \frac{\zeta}{\xi} ( h_{, \nu}^{\nu \mu} - \frac{1}{2} h_{\nu}^{\nu , \mu} )\varphi \partial_{\mu} \phi -\frac{\kappa}{2} \left( \frac{1}{2} \eta^{\mu \nu} h_{\alpha}^{\alpha}  - h^{\mu \nu} \right)( \partial_{\mu} \varphi \partial_{\nu} \phi + \partial_{\mu} \phi \partial_{\nu} \varphi ) 
\end{equation}
leads to the vertex $ \phi  \varphi h $, which, in momentum space, reads 
\begin{equation}\label{eq:Vpph}
    \vcenter{\hbox{\includegraphics[scale=0.6]{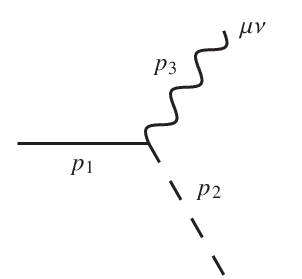}}} \qquad  
\frac{1}{2} i \kappa  \left[\eta^{\mu \nu } \\
   {{p}}_1\cdot
   {p_2}
   - 
    {p_2}^{\mu } {{p}}_1^{\nu
   }-   {{p}}_1^{\mu }
   {p_2}^{\nu }+
   \frac{ \zeta }{\xi}
   (\eta^{\mu \nu }
   {{p}}_1\cdot
   {p_3}- 
   {p_3}^{\mu } {{p}}_1^{\nu
   }-  {{p}}_1^{\mu }
p_3^{\nu } )
\right].
\end{equation}

Similarly, we obtain other relevant vertices for the computation of the counterterm Lagrangian at one-loop order:
\begin{align}\label{eq:ppbpbp}
    & \vcenter{\hbox{\includegraphics[scale=0.70]{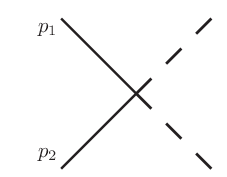}}}
    && 
    \frac{2i\zeta^{2}}{\xi} \kappa^{2} {{p}}_{1 \mu} {{p}}_{2}^{\mu},
\\
    \label{eq:quartic_vertex}
    &    \vcenter{\hbox{\includegraphics[scale=0.70]{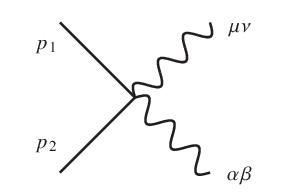}}} 
    &&    
        \begin{aligned}   -i \frac{\kappa^{2}}{4} 
        \big[ & 
 {{p}}_2^{\mu }
   {{p}}_1^{\nu } \eta^{\alpha \beta }+
    {{p}}_1^{\mu }
   {{p}}_2^{\nu } \eta^{\alpha \beta }-
    {{p}}_2^{\beta }
   {{p}}_1^{\nu } \eta^{\alpha \mu }-
    {{p}}_1^{\beta }
        {{p}}_2^{\nu } \eta^{\alpha \mu } \\ & -
    {{p}}_2^{\beta }
   {{p}}_1^{\mu } \eta^{\alpha \nu }-
    {{p}}_1^{\beta }
   {{p}}_2^{\mu } \eta^{\alpha \nu }-
    {{p}}_2^{\alpha }
   {{p}}_1^{\nu } \eta^{\beta \mu }-
    {{p}}_1^{\alpha }
        {{p}}_2^{\nu } \eta^{\beta \mu }\\ & -
    {{p}}_2^{\alpha }
   {{p}}_1^{\mu } \eta^{\beta \nu }-
    {{p}}_1^{\alpha }
   {{p}}_2^{\mu } \eta^{\beta \nu }+
    {{p}}_2^{\alpha }
   {{p}}_1^{\beta } \eta^{\mu \nu }+
    {{p}}_1^{\alpha }
        {{p}}_2^{\beta } \eta^{\mu \nu }\\ & +
    \eta^{\alpha \nu } \eta^{\beta \mu }
   {{p}}_1\cdot
   {{p}}_2+
   \eta^{\alpha \mu } \eta^{\beta \nu }
   {{p}}_1\cdot
   {{p}}_2-
   \eta^{\alpha \beta } \eta^{\mu \nu }
   {{p}}_1\cdot
   {{p}}_2\big],
    \end{aligned}
    \intertext{and the ghost vertex:}
    \label{eq:ghostvertex}
    & \vcenter{\hbox{\includegraphics[scale=0.70]{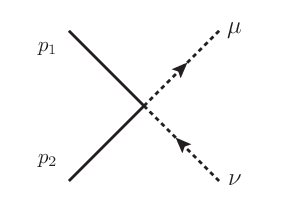}}} 
    &&     
i \zeta \kappa^{2} ({{p}}_{1 }^{\mu}  {{p}}_{2 }^{\nu} + {{p}}_{1}^{\nu} {{p}}_{2}^{\mu} ). 
\end{align}

\section{One-Loop counter-terms} \label{sec:OneLoop}
Here, we compute the diagrams in Figs.~\ref{fig:2point}, \ref{fig:3point} and \ref{fig:4point} in a general background gauge
using the Feynman rules derived in the Appendix~\ref{sec:FR}. We will use dimensional regularization with $ D = 4 -2 \epsilon $, where $D$ is the spacetime dimension.
With these results, we derive the corresponding one-loop counter-term Lagrangians. 

\subsection{$2$-point diagrams}
Here, we present the divergent part of the diagrams shown in Fig.~\ref{fig:2point}. We introduce the following tensor basis:
\begin{subequations} \label{eq:basis}\allowdisplaybreaks
\begin{align} 
    \mathcal{T}^{\mu\nu\alpha\beta}_{(1)} ={}&        k^{\mu} k^{\alpha} k^{\nu} k^{\beta},\\
    \mathcal{T}^{\mu\nu\alpha\beta}_{(2)} ={}& \eta^{\mu\nu}\eta^{\alpha\beta},\\
    \mathcal{T}^{\mu\nu\alpha\beta}_{(3)} ={}& \eta^{\mu\beta} \eta^{\alpha\nu}+\eta^{\mu\alpha}\eta^{\nu\beta},\\
    \mathcal{T}^{\mu\nu\alpha\beta}_{(4)} ={}& k^{\alpha} k^{\beta} \eta^{\mu\nu}+k^{\mu}k^{\nu} \eta^{\alpha\beta},\\
    \mathcal{T}^{\mu\nu\alpha\beta}_{(5)} ={}& k^{\nu} k^{\beta}\eta^{\mu\alpha}+k^{\alpha} k^{\nu} \eta^{\mu\beta}+k^{\mu}k^{\beta} \eta^{\alpha\nu}+k^{\mu} k^{\alpha} \eta^{\nu\beta}.
\end{align}
\end{subequations}

\subsubsection{Graviton loop}
First, we compute the graviton loop, which corresponds to the graphs depicted in Fig.~\ref{fig:2point}(a, b). As result, we obtained the same result found in \cite{Kallosh:1978wt, Barvinsky:1985an, Brandt:2022und} for a general background gauge $ \xi $.
These diagrams lead to the following counter-terms
\begin{equation}\label{eq:VH_grav}
    \mathcal{L}_{\text{2a+2b}} =  \frac{\sqrt{{g}} }{16 \pi^{2} \epsilon} \left [ \left(\frac{1}{120} + \frac{(\xi -1)^2}{6}\right) R^{2}   + \left(\frac{7}{20} + \frac{\xi (\xi -1 )}{3} \right) {R}_{\mu \nu}^{2}\right ].
\end{equation}

\subsubsection{Scalar loop}
Next, we consider the scalar loop diagram shown in Fig.~\ref{fig:2point}(c). 
In the basis \eqref{eq:basis}, we obtain that the scalar loop diagram is equal to 
\begin{equation}\label{eq:scalarloopres}
    \frac{\kappa ^2}{16 \pi^{2} \epsilon }
    \left( \frac{1}{60}\mathcal{T}_1^{\mu \nu \alpha \beta}  +  \frac{1}{80}\mathcal{T}_2^{\mu \nu \alpha \beta}      k^4 + \frac{1}{480}\mathcal{T}_3^{\mu \nu \alpha \beta}   k^4 -\frac{1}{80}\mathcal{T}_4^{\mu \nu \alpha \beta}   k^2-\frac{1}{480}\mathcal{T}_5^{\mu \nu \alpha \beta} k^2.
   \right)
\end{equation}
This result leads to the following counterterms
\begin{equation}\label{eq:VH_scalar_lag}
    \mathcal{L}_{\text{2c}} =  \frac{\sqrt{{g}} }{16 \pi^{2} \epsilon} \left ( \frac{1}{120} {R}_{\mu \nu}^{2} + \frac{1}{240} {R}^{2}\right ).
\end{equation}

\subsubsection{Graviton self-energy}
One can find the graviton self-energy in a general gauge with a scalar field by summing the diagrams in Fig.~\ref{fig:2point}(a, b, c).
The divergent part of the graviton self-energy leads to the counterterm Lagrangian:
\begin{equation}\label{eq:total2}
    \mathcal{L}_{\text{2a+2b+2c}} =  \frac{\sqrt{{g}} }{16 \pi^{2} \epsilon} \left [ 
 \left (   \frac{1}{80} + \frac{1}{6} ( \xi -1)^{2} \right){R}^{2}
 + 
   \left (  \frac{43}{120} + \frac{\xi (\xi -1)}{3}\right){R}_{\mu \nu}^{2}
 \right ],
\end{equation}
which is the result found in \cite{tHooft:1974toh, *thooft:1993}.

\subsubsection{Scalar self-energy} 
The scalar self-energy can be computed in a general gauge $ \xi $. The diagram is shown in Fig.~\ref{fig:2point}(d). 
The divergent part of this diagram reads 
\begin{equation}\label{eq:scalarSE}
   \kappa^{2}  \frac{\zeta }{16 \pi^{2} \epsilon} k^{4}.
\end{equation}
Using the prescription, 
\begin{equation}\label{eq:2mto2s}
    k^{4} \to  \frac{1}{2! } (\mathsf{D}_{\mu} \mathsf{D}^{\mu} \phi )^{2} ,
\end{equation}
one finds the counterterms
\begin{equation}\label{eq:parte2d}
    \mathcal{L}_{\text{2d}} = 
    \frac{ \kappa^{2}\zeta }{2} \frac{\sqrt{\bar{g}} }{16 \pi^{2} \epsilon} (\mathsf{D}_{\mu} \mathsf{D}^{\mu} \phi  )^{2},
\end{equation}
which correspond to one quarter of the 't Hooft and Veltman result (when $ \xi =\zeta =1$). This result also agrees with Ref.~\cite{Grisaru:1975ei} for $ \xi =1$.

\subsection{$3$-point function}\label{sec:3point}

The one-loop diagrams that contributes to the $3$-point function $ \mathfrak{g}_{\mu \nu} (k) {\phi} (k_1) {\phi} (k_{2} )$ are shown in Fig.~\ref{fig:3point}. In order to simplify our results, we will introduce the tensor basis: 
\begin{subequations}\label{eq:tensorbasis2}\allowdisplaybreaks
\begin{align}
    \mathcal{T}^{\mu \nu}_{R} (k,k_{1} , k_{2} )={}&
   -2 k_{1} \cdot k_{2 } ( k^{\mu} k^{\nu} - k^{2} \eta^{\mu \nu} ), \\\nonumber 
    \mathcal{T}^{\mu \nu}_{R_{ \alpha \beta } }(k,k_{1} , k_{2} ) ={}& 
\left({k}\cdot {k}_1\right) \left({k}\cdot {k}_2\right)
\eta^{\mu \nu } + \frac{k^{2}}{2} ( {k}_2{}^{\mu } {k}_1{}^{\nu
   }+  {k}_1{}^{\mu } {k}_2{}^{\nu })\\ & -\frac{1}{2} \big[
   {k}_2{}^{\mu } {k}^{\nu } \left({k}\cdot
   {k}_1\right)+ {k}^{\mu } {k}_2{}^{\nu }
   \left({k}\cdot {k}_1\right)+{k}_1{}^{\mu }
   {k}^{\nu } \left({k}\cdot {k}_2\right)+
   {k}^{\mu } {k}_1{}^{\nu } \left({k}\cdot {k}_2\right)
\big]
   .
\end{align}
\end{subequations}
The tensor structures above corresponds, in momentum space to the invariants: $ R \partial_{\mu} \phi (k_{1} ) \partial^{\mu} \phi ( k_{2} )$ and $ R_{\mu \nu} (k) \partial^{\mu} \phi ( k_{1} ) \partial^{\nu} \phi (k_{2} )$, respectively.

Using this basis, one can show that the divergent part of the diagrams Fig.~\ref{fig:3point}(a,b, c) are respectively given by:
    \begin{align}\label{eq:3pAa}
& 
-\frac{1}{3} \kappa ^3 (\xi -1) (\xi +4) \left [ \frac{1}{4} \mathcal{T}_{R}^{\mu \nu} (k, k_{1} , k_{2} )- \mathcal{T}_{R_{\alpha \beta }}^{\mu \nu} (k, k_{1} , k_{2} )\right ]  
,\\
& +\frac{1}{12 \xi } \kappa ^3   \zeta^{2} \mathcal{T}_{R}^{\mu \nu} (k, k_{1} , k_{2} ),
\\
&
- \kappa ^3 \zeta  \left [ \frac{1}{6} \mathcal{T}_{R}^{\mu \nu} (k, k_{1} , k_{2} )+\mathcal{T}_{R_{\alpha \beta }}^{\mu \nu} (k, k_{1} , k_{2} )\right ].
    \end{align}
These results can be rewritten in position space as:
    \begin{align}\label{eq:3pointINV}
& 
- \frac{1}{3}  \kappa ^2 (\xi -1) (\xi +4) \left[\frac{1}{4} R \partial^{\mu} \phi \partial_{\mu} \phi -   R_{\mu \nu} \partial^{\mu} \phi \partial^{\nu} \phi)\right], 
\\
&+ \frac{1}{12 \xi } \kappa^{2} \zeta^{2} R \partial_{\mu} \phi \partial^{\mu} \phi  
\intertext{and}
&     -\frac{1}{6}  \kappa^{2} \zeta  R \partial^{\mu} \phi \partial_{\mu} \phi -  \kappa^{2} \zeta R_{\mu \nu} \partial^{\mu} \phi \partial^{\nu} \phi.
    \end{align}

Now, we consider the divergent part of the diagrams in Fig.~\ref{fig:3point}(d, e, f). Summing these diagrams, we obtain a transverse result that can be written in terms of the basis \eqref{eq:tensorbasis2}: 
\begin{equation}\label{eq:sum3def}
    \kappa^{3} \left [ \frac{( \xi - \zeta )^{2}}{\xi} \frac{3 \xi -1   }{12 } \mathcal{T}^{\mu \nu}_{R} ( k_{1} + k_{2} , k_{1} , k_{2} ) 
        -\frac{\xi -1 -2 \zeta  }{2} \mathcal{T}^{\mu \nu}_{R_{\alpha \beta } }  
( k_{1} + k_{2} , k_{1} , k_{2} ) 
    \right ].
\end{equation}

The total result, using Eqs.~\eqref{eq:3pointINV} and \eqref{eq:sum3def}, leads to the counterterms: 
\begin{equation}\label{eq:3pointCT}
    \mathcal{L}_{\text{3P}}  = \kappa^{2} \frac{ \sqrt{{g} } }{16 \pi^{2} \epsilon}  \left [ \frac{2 ( \xi -1)^{2} -6 \zeta ( \xi -1 )+2+3 \zeta ( \zeta -2)   }{12} R \partial_{\mu} \phi \partial^{\mu} \phi  
        +\frac{\xi -1}{6} (2 \xi +5)
        R_{\mu \nu} \partial^{\mu} \phi \partial^{\nu} \phi
    \right ].
\end{equation}
When $ \xi = \zeta =1$, it reduces to the result found in \cite{tHooft:1974toh, *thooft:1993}. When $ \xi=1$, we find that 
\begin{equation}\label{eq:3pointGrisaru}
    \mathcal{L}_{\text{3P}}|_{\xi =1}  =    \kappa^{2} \frac{ \sqrt{g}  }{16 \pi^{2} \epsilon }\left[ - \frac{1}{12} + \frac{( \zeta -1)^{2} }{4}\right] R \partial_{\mu} \phi \partial^{\mu} \phi,
\end{equation}
which is the same result found in Ref.~\cite{Grisaru:1975ei}.

\subsection{$4$-point scalar function}
Next, let us consider the divergent part of the $4$-point scalar function. We remember the reader that permutations are implied.

The diagram shown in Fig.~\ref{fig:4point}(a) (summing all the permutations)  
leads to
\begin{equation}\label{eq:4point}
    2 (\xi^{2} + \xi + 1) \kappa ^4 \left[\left({k_1}\cdot
   {k_4}\right)
   \left({k_2}\cdot
   {k_3}\right)+\left({k_1}\cdot {k_3}\right)
   \left({k_2}\cdot
   {k_4}\right)+\left({k_1}\cdot {k_2}\right)
   \left({k_3}\cdot
   {k_4}\right)\right].
\end{equation}
Computing the diagram in Fig.~\ref{fig:4point}(b), we find 
\begin{equation}\label{eq:4pointb} 
    \frac{2 \zeta^{4} }{ \xi^{2}} \kappa ^4 \left[\left({k_1}\cdot
   {k_4}\right)
   \left({k_2}\cdot
   {k_3}\right)+\left({k_1}\cdot {k_3}\right)
   \left({k_2}\cdot
   {k_4}\right)+\left({k_1}\cdot {k_2}\right)
   \left({k_3}\cdot
   {k_4}\right)\right].
\end{equation}
The ghost diagram in Fig.~\ref{fig:4point}(c) yields 
\begin{equation}\label{eq:4pointc}
    -4 \zeta^{2} \kappa ^4 \left[\left({k_1}\cdot
   {k_4}\right)
   \left({k_2}\cdot
   {k_3}\right)+\left({k_1}\cdot {k_3}\right)
   \left({k_2}\cdot
   {k_4}\right)+\left({k_1}\cdot {k_2}\right)
   \left({k_3}\cdot
   {k_4}\right)\right].
\end{equation}

Now, we compute the triangle and box diagram shown in Fig.~\ref{fig:4point}(d, e, f). We find respectively that 
\begin{align}\label{eq:4d}
    -  & 3  ( \xi - \zeta )^{2}  
    \kappa ^4 \left[\left({k_1}\cdot
   {k_4}\right)
   \left({k_2}\cdot
   {k_3}\right)+\left({k_1}\cdot {k_3}\right)
   \left({k_2}\cdot
   {k_4}\right)+\left({k_1}\cdot {k_2}\right)
   \left({k_3}\cdot
{k_4}\right)\right], 
\\
    - & 4  \frac{\zeta^{2}}{ \xi^{2} }   ( \xi - \zeta )^{2}  
    \kappa ^4 
    \left[\left({k_1}\cdot
   {k_4}\right)
   \left({k_2}\cdot
   {k_3}\right)+\left({k_1}\cdot {k_3}\right)
   \left({k_2}\cdot
   {k_4}\right)+\left({k_1}\cdot {k_2}\right)
   \left({k_3}\cdot
{k_4}\right)\right]
\intertext{and}
                                  &  \frac{2}{\xi^2}( \xi - \zeta )^{4}  \kappa ^4 \left[\left({k_1}\cdot
   {k_4}\right)
   \left({k_2}\cdot
   {k_3}\right)+\left({k_1}\cdot {k_3}\right)
   \left({k_2}\cdot
   {k_4}\right)+\left({k_1}\cdot {k_2}\right)
   \left({k_3}\cdot
   {k_4}\right)\right].
\end{align}

The counterterm Lagrangian in a general gauge  induced by the $4$-point function is given by 
\begin{equation}\label{eq:CT_4point}
    \begin{split}
        \mathcal{L}_{\text{4P}} =\frac{1}{8}  \kappa^{4} 
        [4+2( \xi - 1) + {( \xi - \zeta )}^{2} ] 
        \frac{\sqrt{{g}}}{16 \pi^{2} \epsilon}   {( \partial_{\mu} \phi g^{\mu \nu} \partial_\nu \phi )}^{2}
\end{split}.
\end{equation}
When $ \xi = 1 $, one finds the result obtained in Ref.~\cite{Grisaru:1975ei}, now setting $ \zeta =1$, it reduces to the result in t' Hooft and Veltman \cite{tHooft:1974toh, *thooft:1993}. 

\section{Counterterms in general scalar-gravity models} \label{sec:comparison}

In Refs.~\cite{Barvinsky:1993zg, Kamenshchik:2014waa}, a more general scalar-gravity model is considered. It is interesting both for quantum and inflationary cosmology. The action of this improved gravitational theory reads
\begin{equation} \label{eq:cosmo}
    S=\int_{\mathcal{M}} d^{4}x \sqrt{ \bar{g} }
\left(
U( \phi )R
-\frac{1}{2} G ( \phi ) \bar{g}^{\mu\nu}\partial_{\mu}\bar{\phi}\partial_{\nu}\bar{\phi}
- V ( \phi )
\right),
\end{equation}
In principle, the field dependent couplings $U(\phi)$, $G(\phi)$ and $V(\phi)$ are general. 

A general calculation showing the quantum inequivalence of the Jordan and Einstein frames are provided in \cite{Kamenshchik:2014waa}.
Meanwhile, in \cite{Barvinsky:1993zg}, the one-loop counterterms in the De Donder gauge ($ \xi =1$ and $ \zeta =0$) are derived with 
$ U = 1/ \kappa^{2} $ (or equivalently $ U =-1/\kappa^2$ and $R \to - R$) and $G=1$. The counterterms Lagrangian, in our notation, reads
\begin{equation}\label{eq:BK}
    \begin{split}
        \frac{\sqrt{g} }{32\pi^{2} \epsilon }
& 
\Bigg[
\frac{43}{60}\,{R}_{\alpha\beta}^{2}
+\frac{1}{40}\,{R}^{2}
+\frac{5}{4}\kappa^{4}
\left(g^{\alpha\beta}\phi_{,\alpha}\phi_{,\beta}\right)^{2}
+ \kappa^{2}g^{\alpha\beta}\phi_{,\alpha}\phi_{,\beta}\left(
-\frac{1}{3}{R}
+ \kappa^{2}{V}
-2\frac{\partial^{2}{V}}{\partial\phi^{2}}
\right)
\\ & 
+ {R}
\left(
-\frac{13}{3}\kappa^{2}{V}
-\frac{1}{6}\frac{\partial^{2}{V}}{\partial\phi^{2}}
\right)
+\frac{5}{2}\kappa^{4}{V}^{2}
-2\kappa^{2}\left(\frac{\partial{V}}{\partial\phi}\right)^{2}
+\frac{1}{2}\left(\frac{\partial^{2}{V}}{\partial\phi^{2}}\right)^{2}
\Bigg].
\end{split}
\end{equation}
For our simple scalar-gravity model \eqref{eq:21}, we have to set $R \to -R$, $V=0$ in Eq.~\eqref{eq:BK} which leads to Eq.~\eqref{eq:CTLAG}.

\section{On-shell counterterm Lagrangian }\label{sec:onshell}

The counterterm Lagrangian of quantum gravity and scalar matter fields in a general gauge is obtained by adding all the contributions coming from the $2$, $3$ and $4$-point function obtained in Eqs.~\eqref{eq:VH_grav}, \eqref{eq:VH_scalar_lag}, \eqref{eq:parte2d}, \eqref{eq:3pointCT} and \eqref{eq:CT_4point}. This Lagrangian reads (the explicit form is shown Eq.~\eqref{eq:CTLAG}.)
\begin{equation}\label{eq:totallagCT}
    \mathcal{L}_{\text{CT}} = \mathcal{L}_{\text{2P} } + \mathcal{L}_{\text{3P}} + \mathcal{L}_{\text{4P}}, 
\end{equation}
where $ \mathcal{L}_{\text{2P}} = \mathcal{L}_{2a+2b} + \mathcal{L}_{2c} + \mathcal{L}_{2d} $.

Using the on-shell condition \eqref{eq:311}, we have that
\begin{equation}\label{eq:onshell}
    R_{\mu \nu} \partial^{\mu} \phi \partial^{\nu} \phi= R (\partial^{\mu} \phi)^2, \quad R^{2} = R_{\mu \nu}^{2} = - \frac{\kappa^{2}}{2} R (\partial_{\mu} \phi )^{2} \quad \text{and} \quad \kappa^2 \partial_\mu\phi \partial^\mu \phi = -2 R (\partial_\mu \phi)^2.
\end{equation}
To obtain the on-shell counterterm Lagrangian, we consider each term of Eq.~\eqref{eq:totallagCT}:
\begin{subequations}
\begin{align}
\mathcal{L}_{\text{2P}}|_{\text{on-shell}} ={}& - \frac{ \kappa^{2}\sqrt{g}  }{16 \pi^{2} \epsilon }\left[\frac{43}{160} + \frac{\xi^2}{4} - \frac{\xi}{3}\right]R \partial_{\mu} \phi \partial^{\mu} \phi, \\
\mathcal{L}_{\text{3P}}|_{\text{on-shell}} ={}&  \frac{\kappa^{2} \sqrt{g}  }{16 \pi^{2} \epsilon }\left[ -\frac{1}{2} +\frac{\xi^2}{2} + \frac{\xi}{6} + \frac{\zeta^2 -2\xi\zeta}{4}  \right]R \partial_{\mu} \phi \partial^{\mu} \phi, \\
\mathcal{L}_{\text{4P}}|_{\text{on-shell}} ={}& - \frac{ \kappa^{2}\sqrt{g}  }{16 \pi^{2} \epsilon }\left[\frac{1}{2}+\frac{\xi^2}{4}+\frac{ \xi}{2}  + \frac{ \zeta^2-2 \xi \zeta }{4} \right]R \partial_{\mu} \phi \partial^{\mu} \phi. 
\end{align}
\end{subequations}
Remember that $ \mathsf{D}_\mu \mathsf{D}^\mu \phi =0$, this implies that $ \mathcal{L}_{2d}$ vanish on-shell.

Clearly, summing all contribution in the left-hand side of Eq.~\eqref{eq:totallagCT}, the dependence on $\xi$ and $\zeta$ vanish. Then, we arive to the well-know result:
\begin{equation}\label{eq:finalresult}
\begin{split}
    \mathcal{L}_{\text{CT}}|_{\text{on-shell}} ={}& -\frac{\sqrt{g} \kappa^{2}}{16 \pi^{2} \epsilon}\, \frac{203}{160} \ R \partial_{\mu} \phi \partial^{\mu} \phi \\ ={}&\frac{\sqrt{g} \kappa^{4}}{16 \pi^{2} \epsilon}\, \frac{203}{320} \ 
    [ \partial_{\mu} \phi \partial^{\mu} \phi ]^{2},
    \end{split}
\end{equation}
where we have used \eqref{eq:onshell}.

\section{Landau-DeWitt background gauge}\label{sec:appD}

We consider here, for simplicity, the pure quantum gravity, where this gauge is characterized by the condition 
\begin{equation}\label{eq:D1}
    h^{\mu \nu}_{; \nu} - \frac{1}{2} h_{\alpha}^{\alpha ; \mu} =0
\end{equation}
where $ h_{\alpha \beta} = \bar{g}_{\alpha \beta} - g_{\alpha \beta}= \bar{g}_{\alpha \beta} - \eta_{\alpha \beta} - \kappa \mathfrak{g}_{\alpha \beta}  $.
In order to quantise this theory in the Landau-DeWitt gauge, we introduce external sources (antifields) $ G^{* \alpha \beta} $ and $C^{* \mu} $ coupled respectively to the variations of the graviton field $g_{\alpha \beta} $ and the ghost field $ \eta_{\mu} $, under the BRST transformations \eqref{eq:210}.

The full Lagrangian is obtained by the usual BRST procedure 
\begin{equation}\label{eq:D3}
    \mathcal{L}^{\text{full}} =\frac{\sqrt{\bar{g}}}{\kappa^{2}} \bar{R} - \mathsf{s} 
    \left [ - \sqrt{g} \bar{\eta}_{\mu} \left( h^{\mu \nu}_{; \nu} - \frac{1}{2} h_{\alpha}^{\alpha}{}^{; \mu} \right) + \frac{1}{\kappa} G^{* \alpha \beta} \bar{g}_{\alpha \beta} - C^{* \mu} \eta_{\mu} \right ],
\end{equation}
where we have used Eq.~\eqref{eq:211} in the limit $ \xi \to 0$, and omitted  the matter fields.

When applying the BRST operator $ \mathsf{s} $ to the background graviton field $ \mathfrak{g}_{\alpha \beta} $ in the Eq.~\eqref{eq:D3},  it is useful to introduce a source  $ W_{\alpha \beta} $ defined as 
\begin{equation}\label{eq:D4}
    \mathsf{s} \mathfrak{g}_{\alpha \beta} = W_{\alpha \beta}.
\end{equation}
We note that the mass dimension of $ G^{* \alpha \beta} $, $C^{* \mu} $ and $ W_{\alpha \beta } $ is $2$, while their ghost number is respectively $-1, -2, 1$. These sources are invariant under BRST transformations, so that 
\begin{equation}\label{eq:D5}
    \mathsf{s}{G}^{* \alpha \beta} = \mathsf{s} C^{* \mu} = \mathsf{s} W_{\alpha \beta} =0.
\end{equation}

Using the BRST symmetry of the theory, one obtains the Slavnov-Taylor identity for the action $ S = \int \mathop{d^{4} x} \mathcal{L}^{\text{full}} $: 
\begin{equation}\label{eq:D6}
    \begin{split}
        \int & \mathop{d^{4} x} \left [ 
    \frac{\delta S}{\delta g_{\mu \nu}} 
    (\mathsf{s} g_{\mu \nu} ) 
+  
\frac{\delta S}{\delta \eta_{\mu}} 
( \mathsf{s} \eta_{\mu} ) 
+  
\frac{\delta S}{\delta \bar{\eta}_{\mu}} 
( \mathsf{s} \bar{\eta}_{\mu} ) 
+ 
\frac{\delta S}{\delta \mathfrak{h}_{\mu \nu}}
( \mathsf{s} \mathfrak{h}_{\mu \nu} ) 
\right ]     
\\
    & =\int \mathop{d^{4} x} \left [ 
    \frac{\delta S}{\delta g_{\mu \nu}} 
    \frac{\delta S}{\delta G^{* \mu \nu}}  
    + 
\frac{\delta S}{\delta \eta_{\mu}} 
\frac{\delta S}{\delta C^{* \mu}}  
- 
\frac{\delta S}{\delta \bar{\eta}_{\mu}} 
B_{\mu}  
+ 
\frac{\delta S}{\delta \mathfrak{h}_{\mu \nu}} 
W_{\mu \nu} 
\right ] =0.
    \end{split}
\end{equation}
Moreover, using the Eq.~\eqref{eq:28} in the limit $ \xi \to 0$ and $ \zeta \to 0$, one deduces the Nakanishi identity 
\begin{equation}\label{eq:D7}
    \frac{\delta S}{\delta B_{\mu}} = h^{\mu \nu}_{; \nu} - \frac{1}{2} h_{\alpha}^{\alpha ; \mu} .    
\end{equation}

Employing the above  relations, extended to  the quantum effective action $ \Gamma $, one can derive the Ward identities which relate the one-particle irreducible Green functions. We will not go here
into the details of this derivation, which involves rather lengthy calculations. For our purpose, it is sufficient to point out that one of such Ward identities relates the ghost-background graviton-
ghost vertex $ \Gamma_{\eta \bar{\eta} \mathfrak{g}} $ to the  1PI Green function $ \Gamma_{W G^{*} \mathfrak{g}} $. In momentum space, one may represent  schematically this relation  as 
\begin{equation}\label{eq:D8}
    \Gamma_{\eta_{\mu} \bar{\eta}^{\nu} \mathfrak{g}_{\alpha \beta} (k_1,k_2) } \sim k^{\rho} k_{1\sigma} \Gamma_{W_{\mu \rho } G^{* \nu \sigma } \mathfrak{g}_{\alpha \beta} (k_1,k_2)}.   
\end{equation}

We note that in the Yang-Mills theory, a similar relation involves the ghost-background gluon-ghost vertex $ \Gamma_{\eta \bar{\eta} \bar{A} } $ and   the Green function $ \Gamma_{\Omega A^{*} \bar{A}} $. In this case, by power counting, one can see that this Green function is finite, which implies the finiteness of ghost-background gluon-
ghost vertex in the Landau gauge. On  the other hand, one can see by power counting that in quantum gravity, the Green function $ \Gamma_{W G^{*} \mathfrak{g}} $ is quadratically divergent, which implies that the ghost-background graviton-ghost vertex is ultraviolet divergent in the Landau-DeWitt background gauge. This is confirmed by an explicit evaluation of the diagrams shown in Fig.~\ref{fig:vertex}, which yields a very involved divergent result.

\bibliography{GaugeDependency.bib}      
\end{document}